\documentclass{article}
\usepackage{fullpage}
\usepackage{graphicx}
\usepackage{geometry}
\usepackage{epstopdf}
\usepackage{amsmath}
\usepackage{amssymb}
\usepackage{listings}
\usepackage{color}
\usepackage{mwe}
\usepackage{subfigure}
\usepackage{setspace}
\usepackage{hyperref}
\hypersetup{
	colorlinks=true,
	linkcolor=blue,
	filecolor=magenta,      
	urlcolor=cyan,
}

\definecolor{codegreen}{rgb}{0,0.6,0}
\definecolor{codegray}{rgb}{0.5,0.5,0.5}
\definecolor{codepurple}{rgb}{0.58,0,0.82}
\definecolor{backcolour}{rgb}{0.95,0.95,0.92}

\lstdefinestyle{mystyle}{
	backgroundcolor=\color{backcolour},   
	commentstyle=\color{codegreen},
	keywordstyle=\color{magenta},
	numberstyle=\tiny\color{codegray},
	stringstyle=\color{codepurple},
	basicstyle=\footnotesize,
	breakatwhitespace=false,         
	breaklines=true,                 
	captionpos=b,                    
	keepspaces=true,                 
	numbers=left,                    
	numbersep=5pt,                  
	showspaces=false,                
	showstringspaces=false,
	showtabs=false,                  
	tabsize=2
}

\author{Namrata Gupta, Mangal Kothari\footnote{Department of Aerospace Engineering, Indian Institute of Technology, Kanpur, 
	Uttar Pradesh, India, email: mangal@iitk.ac.in}, and Abhishek\footnote{Department of Aerospace Engineering, Indian Institute of Technology, Kanpur, 
	Uttar Pradesh, India, email: abhish@iitk.ac.in}}
\title{Modeling and Control of Inverted Flight of a Variable-Pitch Quadrotor}
\date{}
\begin{document}
	\maketitle

\begin{abstract}
	This paper carries out the mathematical modeling, simulation, and control law design for a quadrotor with variable-pitch propellers. The use of variable-pitch propeller for thrust variation instead of RPM regulation facilitates generation of negative thrust, thereby augmenting the rate of change of thrust generation amenable for aggressive maneuvering. Blade element theory along with momentum theory is used to estimate propeller thrust and torque essential for formulating equation of motion of the vehicle. The proposed flight dynamics model is used for non-linear control design using dynamic inversion technique, which is then used to stabilize, track reference trajectory, and simulate flip maneuver. The rotor torque is an irrational function of the control input which makes the control design challenging. To address this problem, the control design employs three loops. The outer loop solves the translational dynamics to generate the thrust, pitch angle, and roll angle commands required to track the prescribed trajectory. Using the command generated in the outer loop, the inner loop simplifies the rotational dynamics to provide the desired rate of angular velocities. A control allocation loop is added to address the problem of nonlinearity associated with rotor torque. This is done by introducing the derivative of thrust coefficient as a virtual control input. These virtual inputs determine the derivatives of thrust and body moments, which in turn is used to generate the required thrust and body moments. The concept is validated by showing attitude stabilization in real flight for a variable pitch quadrotor. The performance of the proposed design is shown through simulated results for attitude stabilization and trajectory following. Reverse thrust capability of variable-pitch quadrotor is also shown by performing flip maneuver in which quadrotor roll angle changes from 0 to 180 degrees.
\end{abstract}

\section{Introduction}

Last decade saw the development of various configurations of Unmanned Aerial Vehicles (UAVs) capable of hovering flight. UAV configurations ranging from flapping wing, rotary wing to cycloidal rotor concept have been developed and studied in recent years (see Refs.~\cite{beers,pranay,felipe,mufly,robofly,naudin,moble}). Some of these designs have seen greater success than others. Researchers around the world have been working on different rotary wing configurations such as the Micro Coaxial Rotorcraft (MICOR)~\cite{felipe}, muFly~\cite{mufly}, RoboFly or the Samara Micro Air Vehicle (MAV) -- a prop assisted mono blade~\cite{robofly}, Coanda UAV~\cite{naudin}, and Cycloidal rotor MAV~\cite{moble}. Among these the most successful configuration which caught the eyes of researchers and amateurs alike in the early 2000s is the quadrotor configuration. Since then, the quadrotors have been extensively studied and several papers have been authored studying their dynamics, stabilization, and control. Some of the notable pioneering works include that by Bouabdallah {\em et al.}~\cite{bouab1,bouab2} on ETH Z$\ddot{u}$rich's `OS4' (a belt-driven indoor quadrotor vehicle) and Castillo {\em et al.} ~\cite{castillo}. The commercial success for this configuration can be gauged from the fact that quadrotors with all up weights ranging from 50 grams to 15--20 kg can be bought off-the-shelf and can be used for a variety of missions.

The conventional quadrotor with fixed pitch propellers is controlled by varying the RPM of the individual motors and suffers from a few limitations: i) the rotational inertia of the motors limits the control bandwidth of the system~\cite{cutler0}; and ii) the stabilization of larger quadrotors through RPM control alone becomes challenging as a point can be reached where the torque required to change the RPM of the motor exceeds the capacity of the motor. Due to these limitations, the current flight control strategy of quadrotors is not suitable for larger full scale vehicles meant for lifting heavy payload. 

These limitations can be overcome by employing a quadrotor design with variable-pitch control~\cite{abhish_ahs}. It would appear that the use of variable-pitch propellers add complexity to a simple and robust quadrotor design. But, the advantages of increased controller bandwidth due to the availability of reverse thrust from propellers and scalability to full scale size justify the design. The idea of variable-pitch propeller based quadrotor is an old one. It has somehow not attracted enough attention from researchers, until recently. In 1922, Georges de Bothezat and Ivan Jerome built and flew the ``Flying Octopus'' a quadrotor with rotors located at each end of a truss structure of intersecting beams, placed in the shape of a cross. Control of the machine was achieved by changing the pitch of each of the propellers~\cite{leishman}. The Hoverbot, developed at University of Michigan by Johann Borenstein~\cite{boren} is the first documented effort at designing and flying a small scale quadrotor with variable pitch control. However, the Hoverbot never achieved flight beyond tethered hovering. 

In recent past, while, several hobbyists have demonstrated the construction and flight of remote controlled variable-pitch quadrotors, a serious and organized effort of studying the flight mechanics and control of a variable-pitch quadrotor was demonstrated by Cutler {\em et al.} ~\cite{cutler0,cutler1}. To keep the design simple, the four motor based design of fixed pitch quadrotor was retained and a mechanism similar to tail rotor swashplate was used to change the blade pitch angle for each of the propellers. The flight performance of variable-pitch quadrotor with fixed pitch ones was systematically studied and the following conclusions were made: (i) the variable-pitch propeller quadrotor could generate significantly large rate of change of thrust when compared to a fixed-pitch design, thereby improving the capability to perform aggressive maneuvers; and (ii) the ability to generate negative thrust by variable-pitch propellers can be utilized to perform aerobatics and allow for inverted flight. However, the dynamics of variable-pitch quadrotor UAV has not been studied in detail and therefore model based control has not been applied for control and navigation of such quadrotors until recently. It should be noted that the thrust for variable pitch propellers not only depend on RPM but also on the blade pitch angle. Unlike RPM controlled quadrotors, designing a controller for these vehicles is challenging as the function representing the relation between thrust and torque with blade pitch angle is not rational. Therefore, the control design methods available for RPM controlled quadrotors cannot be applied as it is for variable-pitch quadrotors. Previously, approaches like backstepping~\cite{Madani_quad2006,nagaty2013control}, sliding mode ~\cite{Bouabdallah_quad2005}, nonlinear dynamic inversion (NDI) ~\cite{Das_quad2009},~\cite{prabhumangal}, adaptive control \cite{Zachary_quad2010,Lee_quad2009,mohammadi2013adaptive,shastry2016neuro}, have been applied to design controller for quadrotors. A comparison study was carried out in~\cite{carrillo2012hovering} using visual feedback for stabilization and tracking. 

Cutler et al.~\cite{cutler2015analysis}  carried out the analysis and control design for the variable-pitch quadrotor. However, the analytical model developed for the variable-pitch quadrotor in this paper suffers from some serious errors. First, in the paper the angle of attack used for computing the blade lift and drag forces has erroneously been replaced with geometric blade pitch angle ignoring the contribution of the blade induced velocity (inflow) component which has significant effect on angle of attack. The inflow angle is a function of thrust being generated by the rotor which in turn is a function of blade pitch angle itself. Therefore, ignoring the induced inflow angle would result in significant overestimation of the lift and drag forces. Further, this mistake simplifies the control design.  Second, the paper assumes that the multiplication of a constant ``drag coefficient'' with thrust force would result in yawing moment. This is only valid for a fixed-pitch propeller which has constant thrust and torque coefficients. The torque responsible for yawing motion of a variable pitch quadrotor is a function of blade pitch angle itself and therefore cannot be obtained by merely multiplying the thrust with a constant factor for entire range of pitch angles. While the theoretical model was incorrect, authors in~\cite{cutler2015analysis} were able to demonstrate the flight of variable pitch quadrotor using PID control design. The flight dynamics model of variable-pitch quadrotor aerial vehicle was first proposed in~\cite{namrata} in which the controller was used to perform attitude stabilization and trajectory tracking. The detailed control design and ability  to perform flip maneuver has not been studied for a variable-pitch quadrotor using a model based controller and this is the focus of the present paper. 

The present paper focuses on development of flight dynamics model based on Blade Element Theory (BET) and uniform inflow for variable pitch quadrotor is presented. Next, the control design based on dynamic inversion technique is developed for aggressive maneuvering of variable-pitch quadrotors. The challenge associated with the control allocation in variable-pitch quadrotors is addressed by use of an additional loop that dynamically allocates control to generate the desired thrust and moments. Finally, the nonlinear controller is  used to simulate the stabilization, flip and upright/inverted trajectory following of the variable pitch quadrotor. The contributions of this paper are two folds: (i) establish the detailed dynamics model of variable-pitch quadrotor aerial vehicle; and (ii) develop and apply a nonlinear controller for stabilization, trajectory tracking and inverted flight. In addition to this, the new design is validated by showing attitude stabilization in real flight.


\section{Quadrotor Modeling}\label{sec:quadmodel}

The strategy for controlling a variable-pitch quadrotor is significantly different from that of conventional fixed-pitch propeller based quadrotor and is discussed in this section. After establishing the control strategy, the six degrees of freedom (six-DOF) Newton-Euler equations representing the dynamics of variable-pitch quadrotor vehicle are derived.

\subsection{Strategy for Control}

Like the conventional quadrotor, the primary control of various motions (three translational, roll, and pitch motions) for variable-pitch quadrotor is achieved by changing the thrust of different rotors in various combinations. However, the mechanism of thrust variation is different. The change in thrust is achieved by simultaneously changing the pitch angle (collective angle) of all the blades. The control of yawing motion and the mechanism involved is significantly different as discussed below. It should be noted that any point of time all the rotors are operated at the same nominal RPM which may be regulated about the specified value for setting the baseline value of thrust.

\begin{figure}[ht]
	\centering
	\includegraphics[width = \textwidth]{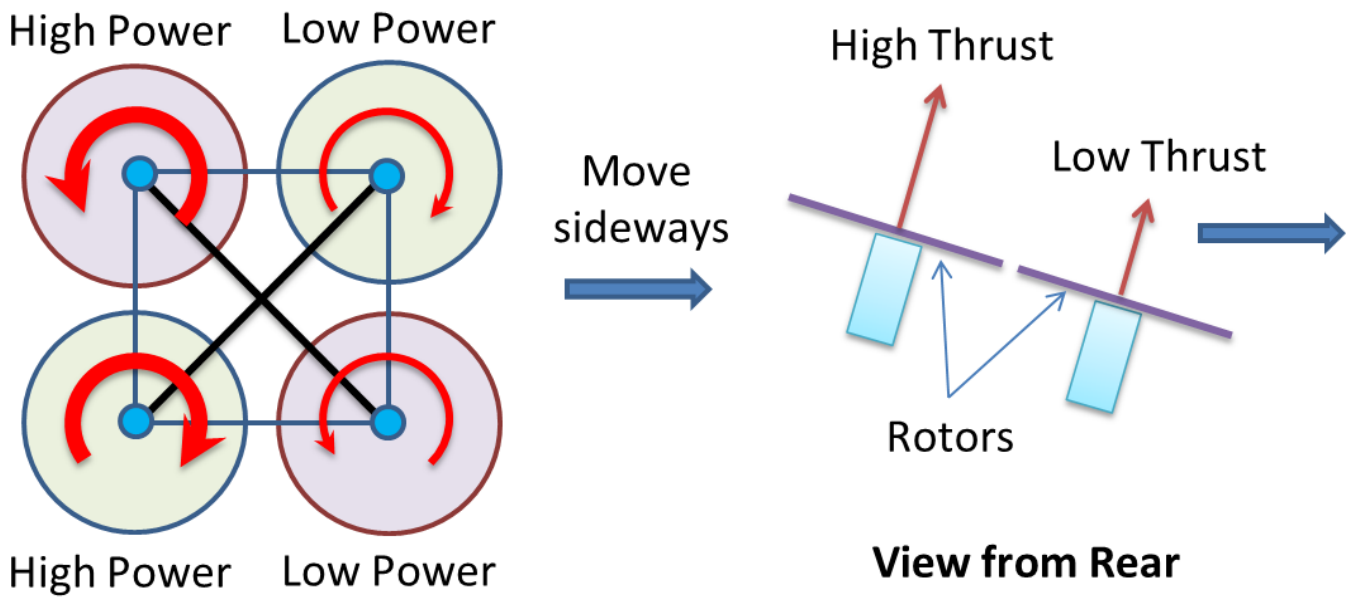}
	\caption{\textbf{Translational flight and roll motion control of variable pitch quadrotor}} 
	\label{quad-con-1}
\end{figure}
The up/down motion is easily controlled by collectively increasing or decreasing the collective pitch angles for all the rotors / propellers simultaneously. Side-wards flight can be achieved as explained in Fig. ~\ref{quad-con-1}. For example, increasing the collective input and thereby increasing the thrust of the two left rotors lifts the left side up and generates a net thrust component to the right. Consequently, the quadrotor would move to the right. The change in torque/power of the two left rotors is equal and opposite, therefore, the moment remains balanced and pure translational motion can be achieved. By the same principle, increasing the collective of the two rear rotors would result in forward flight. The yaw control is less intuitive. 

\begin{figure}[ht]
	\centering
	\includegraphics[width = \textwidth]{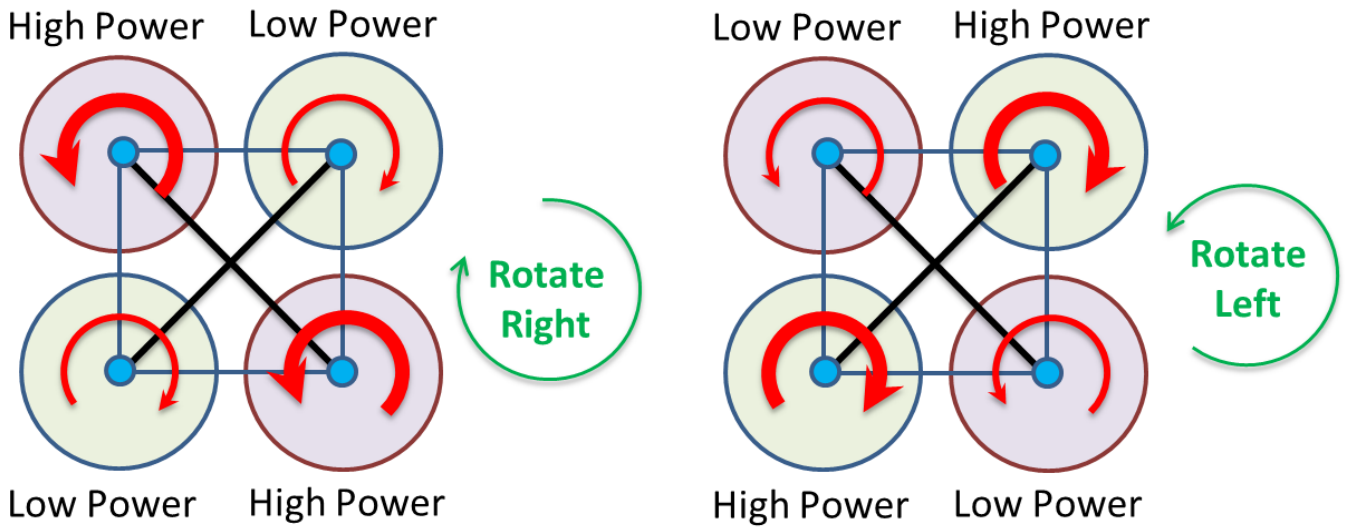}
	\caption{\textbf{Yaw motion control of variable pitch quadrotor}} 
	\label{quad-con-2}
\end{figure}

The method of generating yawing moment is identical to that used for coaxial and tandem helicopters and is known as ``differential collective''. In this, the collective pitch of the two diagonal rotors rotating in the same direction is increased and the collective pitch of the other diagonal pair is reduced. The increased collective pitch results in increasing the lift and drag forces experienced by both these rotors, while the other two rotors would experience an identical reduction in lift. The rotors with increased lift and drag would experience an increase in profile and induced torque components compared to the other two rotors which would experience a decrease in the total torque. This net increase in the combined torque of all the rotors would result in yawing motion of the quadrotor as explained in Fig.~\ref{quad-con-2}. It should be noted that this operation has no effect on translational motion, as the combined thrust of all four rotors remain unchanged.

\begin{figure}[ht]
	\centering
	\includegraphics[width = 0.6\textwidth]{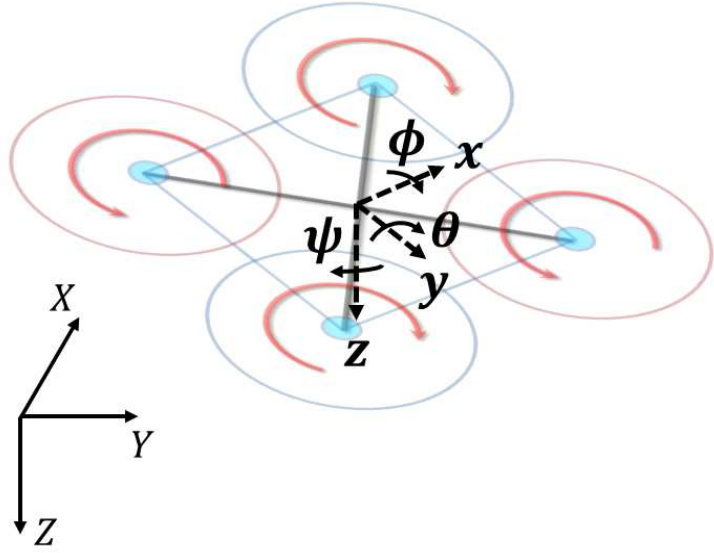}
	\caption{\textbf{Coordinate systems used for development of equation of motion}} 
	\label{coordinate}
\end{figure}

\subsection{Kinematics}

For describing the rigid body dynamics of the quadrotor two coordinate systems, as shown in Fig.~\ref{coordinate}, are employed:  the inertial and body fixed coordinates. All the physical quantities are transformed between the two coordinate systems using the classical Euler angles ($\phi$-roll, $\theta$-pitch, $\psi$-yaw). For modeling using quaternions, refer ~\cite{parwana2017quaternions}. The following expression relates the velocity of quadrotor in these two frames:
\begin{equation}
\begin{split}
\frac{d}{dt} \left(\begin{array}{c}
x\\
y\\
z
\end{array} \right) = R_b^i \left[ \begin{array}{c}
u\\
v\\
w
\end{array} \right] 
\end{split}
\label{6dof_1}
\end{equation}
where
\begin{equation*}
R_b^i = \left[\begin{array}{ccc}
C\theta C\psi & S\phi S\theta C\psi - C\phi S\psi & C\phi S\theta C\psi + S\phi S\psi\\
C\theta S\psi & S\phi S\theta S\psi + C\phi C\psi & C\phi S\theta S\psi -S\phi C\psi\\
-S\theta & S\phi C\theta & C\phi C\theta
\end{array}
\right],
\end{equation*}
$C \beta \triangleq \cos \beta$ and $S \beta \triangleq \sin \beta$. [$x$,$y$,$z$] are the position in the inertial frame and [$u$,$v$,$w$] are the velocity components in the body frame. Similarly, the following expression relates the body rates to Euler angle rates:
\begin{equation}
\left[\begin{array}{c}
\dot{\phi}\\
\dot{\theta}\\
\dot{\psi}
\end{array}\right] = \left[\begin{array}{ccc}
1 & \frac{S\phi S\theta}{C\theta} & \frac{C\phi S\theta}{C\theta}\\
0 & C\phi & -S\phi\\
0 & \frac{S\phi}{C\theta} & \frac{C\phi}{C\theta}\\ 
\end{array}\right]\left[\begin{array}{c}
p\\
q\\
r
\end{array}\right]
\label{rot_trans_inv_sing}
\end{equation}
where [$p$,$q$,$r$] are the angular velocity components (roll, pitch, and yaw) in the body frame.

\subsection{Dynamics}
The rigid body equation of motion of the quadrotor can be derived by applying the linear momentum and angular momentum conservation laws.  In present work, propulsive forces (thrust and torque from the motors) and the gravitational forces are assumed to be the dominant forces. The aerodynamic forces (such as lift and drag) acting on the fuselage are neglected assuming them to be very small. Transforming the gravitational force to the body coordinate axes, the translation dynamics of the quadrotor is given as follows:
\begin{equation} 
\left[\begin{array}{c}
\dot{u}
\\\dot{v}
\\\dot{w}
\end{array}
\right] = 
\left[\begin{array}{c}
0
\\0
\\\frac{T*flag}{M}
\end{array}
\right] + {[R_b^i]}^T
\left[\begin{array}{c}
0\\
0\\
g
\end{array}
\right] +
\left[\begin{array}{c}
rv-qw\\
pw-ur\\
qu-pv
\end{array}
\right]
\label{trans_body}
\end{equation} 
$$flag = -sgn(cos\phi)$$
where $T$ is the total thrust from all the rotors, $M$ is the mass of the quadrotor, and $g$ represents gravitational acceleration. Eq.~(\ref{trans_body}) expresses quadrotor's translational dynamics in the body fixed coordinate system. The variable $flag$ decides the direction of thrust vector in the body coordinate system and is negative for roll angle ($\phi$) less than $90^{\circ}$ and becomes positive causing reversal of thrust direction for $\phi$ greater than $90^{\circ}$ and less than or equal to $180^{\circ}$. The translational dynamics can also be be expressed in the inertial frame as:

\begin{equation}
\left[\begin{array}{c}
\ddot{x}
\\\ddot{y}
\\\ddot{z}
\end{array}
\right] = R_b^i
\left[\begin{array}{c}
0
\\0
\\ -\frac{T}{M}
\end{array}
\right] + 
\left[\begin{array}{c}
0
\\ 0
\\ g
\end{array}
\right] 
\label{6dof_2}
\end{equation} 
Note that the translation dynamics is presented in both the body and inertial frames  for its application in the control design. It is safe to assume that the quadrotor is symmetric about $x$ and $y$ axes, which allows for the rotational dynamics to be represented as:
\begin{equation}
\left[\begin{array}{c}
\dot{p}\\
\dot{q}\\
\dot{r}\\ 
\end{array}\right] = \left[\begin{array}{c}
\frac{I_{yy} - I_{zz}}{I_{xx}} qr\\
\frac{I_{zz} - I_{xx}}{I_{yy}} pr\\
\frac{I_{xx} - I_{yy}}{I_{zz}} pq\\ 
\end{array}\right] + \left[\begin{array}{c}
\frac{l}{I_{xx}}\\
\frac{m}{I_{yy}}\\
\frac{n}{I_{zz}}\\\end{array}\right] 
\label{lmn_equ_invert}
\end{equation}
where  $I_{xx}$, $I_{yy}$, and $I_{zz}$ are moments of inertia about x-axis, y-axis, and, z-axis, respectively.
By the virtue of symmetry, the product of inertia terms are assumed to be zero. Here,  $l$, $m$, and $n$ are the components of the externally applied moments known as rolling, pitching, and yawing moments, respectively. Eqs.~(\ref{6dof_1})-(\ref{lmn_equ_invert}) together represent the complete equation representing the full six degrees of freedom for the quadrotor. 

\subsection{Rotor Dynamics}
Unlike the conventional fixed-pitch quadrotors, the thrust from individual rotors, $T_i$, is varied by changing their collective pitch input. The thrust and moment equilibrium equations for the ``H'' configuration (similar to ``X'' configuration) of the quadrotor is derived about hover condition as shown below. Blade element theory along with momentum theory~\cite{leishman} is  used to calculate thrust and torque of each rotor as a function of thrust coefficient.

With the assumption of the blades being rigid, the aerodynamic forces and moments generated by each rotor can be calculated using blade element theory in which each blade is divided in to a number of elements such that each element is a 2D airfoil. In this, the contribution of each blade element to the total airload (lift, drag, and pitching moment) is calculated and then integrated over the blade radius to calculate the net thrust and torque contribution of each blade, which is then multiplied with number of blades of calculate the total thrust and torque from each rotor. 

Using the approach given in~\cite{leishman}, the non-dimensional thrust coefficient, $C_{T_i}$, and torque coefficient, $C_{Q_i}$, for the i$^{th}$ rotor are given by
\begin{eqnarray}\label{coeff_of_thrust}
C_{T_i} &=&\frac{1}{2}\sigma C_{l_\alpha} \left( \frac{\theta_{0_i}}{3}-\frac{\lambda_i}{2} \right)\\
C_{Q_i} &=& \frac{1}{2}\sigma \left( \frac{\lambda_i C_{l\alpha}\theta_{0_i}}{3}-\frac{\lambda_i ^2 C_{l_\alpha}}{2} + \frac{C_{d_{o_i}}}{4} \right)
\label{coeff_of_torque}
\end{eqnarray}
where $C_{l\alpha}$ is the lift curve slope, $\theta_{0_i}$ is the blade collective pitch angle of the i$^{th}$ rotor,  
$\lambda_i$ is the induced inflow of the i$^{th}$ rotor, $C_{d_{o_i}}$ is the zero lift drag coefficient of the airfoil of the i$^{th}$ rotor, $\sigma=\frac{N_b c}{\pi R}$. Here, $N_b$ is number of blades, $c$ is the chord length of the rotor, $R$ is the rotor blade radius. These non-dimensional quantities can be converted to corresponding dimensional parameters by using $T_i = C_{T_i} \rho A {V_{tip}}^2$ and $Q_i = C_{Q_i} \rho A R {V_{tip}}^2$, where $\rho$ is the density of air, $A$ is the rotor disk area, ${V_{tip}}=\Omega R$ is the tip  speed of rotor blade rotating with angular speed of $\Omega$. The only unknown parameter in Eqs.~(\ref{coeff_of_thrust}) and~(\ref{coeff_of_torque}) is the inflow ratio $\lambda_i$ which can be evaluated using momentum theory for the hovering flight condition and is given by Eq.~(\ref{inflow})
\begin{equation}
\lambda_i= \sqrt{\frac{C_{T_i}}{2}}
\label{inflow}
\end{equation}
Substituting the value of $\lambda_i$ in Eq.~(\ref{coeff_of_thrust}) gives
\begin{equation}
C_{T_i} =\frac{1}{2}\sigma C_{l\alpha}\left(\frac{\theta_{0_i}}{3}-\frac{1}{2}\sqrt{\frac{C_{T_i}}{2}}\right)
\label{modified_coeff_of_thrust}
\end{equation}
which upon rearrangement yields 
\begin{equation}
\theta_{0_i}= \frac{6C_{T_i}}{\sigma C_{l\alpha}}+\frac{3}{2}\sqrt{\frac{C_{T_i}}{2}}
\label{collective}
\end{equation}
Using the above pitch angle and inflow ratio in Eq.~(\ref{coeff_of_torque}) gives
\begin{equation}
C_{Q_i} = \frac{1}{2}\sigma \left(\frac{\sqrt{2}{C_{T_i}}^{\frac{3}{2}}}{\sigma} + \frac{C_{d_{o_i}}}{4}\right)
\label{modified_coeff_of_torque}
\end{equation}
From the above definitions of thrust and torque it can be seen that 
\begin{equation}
T_i = KC_{T_i}
\label{rotor_thrust}
\end{equation}
\begin{equation}
Q_i = KRC_{Q_i}
\label{rotor_torque}
\end{equation}
where $K = \rho A{V_{tip}}^2$. $K$ is typically constant for the variable pitch quadrotor as the rotor speed is regulated about a prescribed constant value. The total thrust generated by the vehicle is then given as:
\begin{align} 
T &= flag\times(T_1 + T_2 + T_3 + T_4) \\
T &=  flag\times K(C_{T_1}+C_{T_2}+C_{T_3}+C_{T_4})
\end{align}
Rolling and pitching moments are obtained by cross multiplying thrust from each rotor with its respective moment arm. Yawing moment is obtained from Eq.~(\ref{modified_coeff_of_torque}). Due to the relative sense of rotation, Rotors 1 and 3 produce torque in positive $z$ direction, while Rotors 2 and 4 produce torque in opposite direction. The contribution of blade drag to total torque (shown in Eq.~(\ref{modified_coeff_of_torque})) is independent of thrust and hence remains constant at all times and cancels out for four rotors. The final expressions for total forces and moments acting on the quadrotor are shown in Eq.~(\ref{force_omega}).
\begin{equation}
\begin{split}
T &=  flag\times K(C_{T_1}+C_{T_2}+C_{T_3}+C_{T_4})\\
l & = d \times K(C_{T_1}-C_{T_2}-C_{T_3}+C_{T_4})\\
m & = -flag\times Kd(C_{T_1}+C_{T_2}-C_{T_3}-C_{T_4})\\
n & = -flag \times \frac{KR}{\sqrt{2}}(|{C_{T_1}|}^{\frac{3}{2}}-|{C_{T_2}}|^{\frac{3}{2}}+|{C_{T_3}}|^{\frac{3}{2}}-|{C_{T_4}}|^{\frac{3}{2}})
\label{force_omega}
\end{split}
\end{equation}
where $d$ is the moment arm of rotors from the center of gravity.

\section{Control Design}\label{sec:controlDesign}

This section develops a controller for variable-pitch quadrotor for stabilization, tracking, and aggressive maneuvers using nonlinear dynamic inversion approach~\cite{DI_Enns},~\cite{prabhumangal}.  The variable-pitch quadrotor is an under actuated system like the conventional RPM regulated quadrotor. However, the dynamics of variable-pitch quadrotor is relatively more complex than the conventional quadrotors as the rotor thrust, roll, and pitch moment equations are linear functions of control input whereas the roll moment is a nonlinear function of control input. Therefore, a closed form solution to these equations is not possible and iterative online solution of the system of equations is tedious and impractical. To address this problem, the control design incorporates three loops: outer loop, inner loop, and control allocation loop. Note that the outer and inner loops are similar to the conventional design. This means that the outer loop is responsible for trajectory tracking whereas the inner loop provides stability. An extra loop is added to dynamically allocate control to determine blade pitch angles of individual rotors. 

Let the state of variable-pitch quadrotor be  $X \triangleq [x~y~z~\phi~\theta~\psi~u~v~w~p~q~r]^T$. For tracking and stabilization, the output of quadrotor is chosen as  $Y \triangleq[x\quad y \quad z\quad \phi\quad \theta\quad \psi]^T$. The control objective is to drive $Y$ to some desired output, $Y_d$. In order to achieve this, the proposed design use a two loop structure  by exploiting the time scale separation principle. The outer loop operates on position $y_{out}=[x \quad y \quad z]^T$ and generates the desired thrust, $T_d$, roll angle, $\phi_d$, and pitch angle,   $\theta_d$. The inner loop drives  $y_{in} =[\phi \quad \theta \quad \psi]^T$ to $y_{{in}_d} =[\phi_d \quad \theta_d \quad \psi_d]^T$ by generating $U_{in} =[l_d \quad m_d \quad n_d]^T$. As the relation between torque and blade pitch angle is nonlinear, a control allocation loop is included to solve the problem of nonlinearities. For this loop, the derivatives of thrust coefficients act as virtual inputs, the value of which needs to be determined to generate the desired thrust and moments.  The thrust coefficients are computed by integrating the derivatives of thrust coefficients. The required blade pitch angle for individual rotors is then calculated. The control allocation loop computes the required blade pitch angles to generate the desired thrust and moments.   

To differentiate whether the quadrotor is upright or inverted, while tracking the given trajectory, a command variable $\sigma_d$ is introduced, where $\sigma_d = sgn(cos\phi_d)$. $\sigma_d$ is negative when  quadrotor is in the inverted flight, i.e, $\phi_d$ is greater than $90^o$. Another variable $flip$, which is set to zero to begin with, is used to check if the quadrotor has achieved required $\phi_d$. The variable $flip$ becomes  1, once desired $\phi_d$ is achieved. 

\subsubsection{Outer Loop Design} The tracking error in position can be defined as $e \overset{\Delta}{=} y_{out} - y_{out_d}$, where $y_{out}$ and $y_{out_d}$ are the current and desired outputs of a quadrotor in the inertial frame. As the relative degree is two, we choose second order stable error dynamics to synthesize the control as follows
\begin{equation}
\ddot{e} + 2\zeta \omega_n \dot{e} + \omega_n^2 ~e = 0
\label{DI_angle_error_dyn}
\end{equation}
From Eq.~\ref{6dof_2} and Eq.~\ref{DI_angle_error_dyn},  we get
\begin{multline}  
R_b^i
\left[\begin{array}{c}
0
\\0
\\ -\frac{T}{M}
\end{array}
\right] + 
\left[\begin{array}{c}
0
\\ 0
\\ g
\end{array}
\right] = 
\left[\begin{array}{c}
\ddot{x}_d
\\\ddot{y}_d
\\\ddot{z}_d
\end{array}
\right] + 2\zeta_{out} {\omega_{n_{out}}}^T
\left[\begin{array}{c}
\dot{x}_d - \dot{x}
\\ \dot{y}_d - \dot{y}
\\ \dot{z}_d - \dot{z}
\end{array}
\right]
+\omega_{n_{out}}{\omega_{n_{out}}}^T
\left[\begin{array}{c}
x_d - x
\\ y_d - y
\\ z_d - z
\end{array}
\right] 
\end{multline} 

$\zeta_{out}$ and $\omega_{n_{out}}$ are $3\times1$ matrices. The required thrust and desired roll and pitch angles are given as:
\begin{equation}
\begin{split}
T_d &= Mflag\sqrt{\ddot{x}^2 + \ddot{y}^2 + (g-\ddot{z})^2}\\
\phi_d &=  \sin^{-1} (u_x  \sin\psi_d - u_y  \cos\psi_d)\\
\theta_d &=  \sin^{-1} \frac{u_x  \cos\psi_d + u_y  \sin\psi_d}{ \cos{\phi_d}}
\end{split}
\label{outer_loop_DI_formulation}
\end{equation}
where
\begin{equation*}
\begin{split}
u_x &= M\ddot{x}/T_d \\
u_y &= M\ddot{y}/T_d 
\end{split}
\end{equation*}
The quadrotor is commanded to perform flip maneuver by setting $\sigma_d$  negative, which commands the quadrotor to flip itself, before tracking the trajectory (this means $flip = 0$). At this stage, controller only tracks the altitude and attitude and hence first and second terms of $\omega_{n_{out}}$ are zero, as a result $\ddot{x}$ and $\ddot{y}$ are zero. The variables $\phi_d$ and $\theta_d$ remain the same. After substituting $\ddot{x}$ and $\ddot{y}$ in Eq.~\ref{outer_loop_DI_formulation}, the desired thrust is given by $T_d = M*flag*abs(\ddot{z}-g)$.
Once the quadrotor is flipped ($flip = 1$), it is commanded to follow the given trajectory and generate acceleration along $x$ and $y$ directions, hence the first and second terms of $\omega_{n_{out}}$ become non-zero. The expressions for $T_d$ and $\theta_d$ remain the same as in Eq.~(\ref{outer_loop_DI_formulation}) whereas $\phi_d$ can be expressed as:
\begin{equation}
\phi_d =  \pi - \sin^{-1} (u_x  \sin\psi_d - u_y  \cos\psi_d)\\
\label{phid_inverted}
\end{equation}

\subsubsection{Inner Loop Design}
For designing the inner loop, we again choose the second order stable error dynamics on attitude as:
\begin{equation}
\left[\begin{array}{c}
\ddot{\phi}
\\\ddot{\theta}
\\\ddot{\psi}
\end{array}
\right] = 
\left[\begin{array}{c}
\ddot{\phi}_d
\\\ddot{\theta}_d
\\\ddot{\psi}_d
\end{array}
\right] + 2\zeta_{in} {\omega_{n_{in}}}^T
\left[\begin{array}{c}
\dot{\phi}_d - \dot{\phi}
\\ \dot{\theta}_d - \dot{\theta}
\\ \dot{\psi}_d - \dot{\psi}
\end{array}
\right] +\omega_{n_{in}}{\omega_{n_{in}}}^T
\left[\begin{array}{c}
\phi_d - \phi
\\ \theta_d - \theta
\\ \psi_d -  \psi
\end{array}
\right]
\label{error_dyn_at_di}
\end{equation} 
The Eulerian angular rates in  Eq.~\ref{error_dyn_at_di} is obtained by transforming the body rates to Eulerian rates using Eq.~\ref{rot_trans_inv_sing}. The onboard sensors measure the rate of rotation of a quadrotor in the body frame. The Eulerian angular acceleration computed using the error dynamics is transformed to obtain desired body angular acceleration as follows
\begin{multline}
\left[\begin{array}{c}
\dot{p}\\
\dot{q}\\
\dot{r}
\end{array}\right] = 
\left[\begin{array}{ccc}
1 & 0 & - S{\theta}\\
0 &  C{\phi} &  S{\phi} C{\theta}\\
0 & - S{\phi} &  C{\phi} C{\theta}
\end{array}\right] 
\left[\begin{array}{c}
\ddot{\phi}\\
\ddot{\theta}\\
\ddot{\psi}
\end{array}
\right]
+\left[\begin{array}{ccc}
0 & 0 &  C{\theta} \dot{\theta}\\
0 &  S{\phi} \dot{\phi} & - S{\phi}  S{\theta} \dot{\theta} -  C{\phi} C{\theta} \dot{\phi} \\
0 &  C{\phi} \dot{\phi} & - S{\theta}  C{\phi} \dot{\theta} +  S{\phi} C{\theta} \dot{\phi}
\end{array}\right]
\left [\begin{array} {c}
\dot{\phi}\\
\dot{\theta}\\
\dot{\psi}
\end{array}
\right]
\label{desired_body_euler_acc}
\end{multline}
This desired body angular acceleration relation in Eq.~\ref{desired_body_euler_acc} is obtained by first inverting the Eq.~\ref{rot_trans_inv_sing} and then differentiating it with respect to time. In order to generate the desired body rates in Eq.~\ref{desired_body_euler_acc}, the quadrotor needs to generate the following moments: 
\begin{equation}
\left[\begin{array}{c}
l_d\\
m_d\\
n_d\\ 
\end{array}\right] =  \left[\begin{array}{c}
I_{xx} \dot{p} + (I_{zz} - I_{yy}) qr\\
I_{yy} \dot{q} + (I_{xx} - I_{zz}) pr\\
I_{zz} \dot{r} + (I_{yy} - I_{xx}) pq\\ 
\end{array}\right]
\label{body_acc_2_moments}
\end{equation}  
Next, the task is to determine the required thrust coefficient (blade pitch angle) to generate the thrust and moments calculated from the outer and inner loops.

\subsubsection{Control Allocation Loop} 
\begin{figure*}[h]
	\centering
	\includegraphics[width = \textwidth]{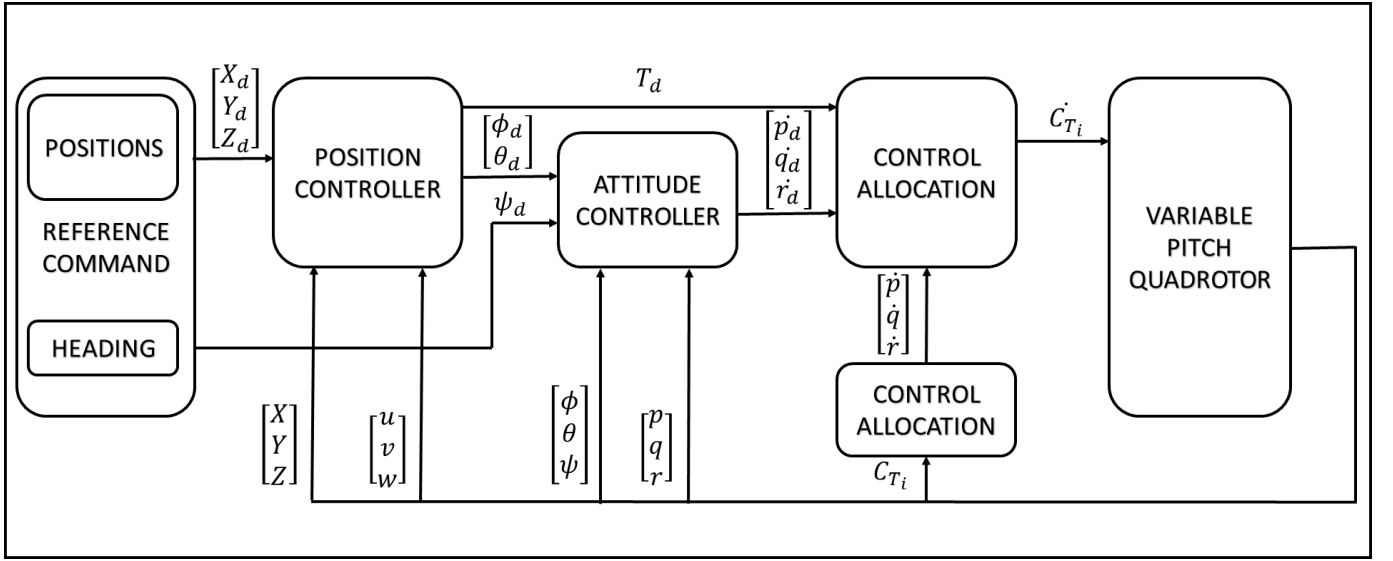}
	\caption{\textbf{A control architecture block}} 
	\label{control-scheme}
\end{figure*}

For the given $T_d$, $l_d$, $m_d$, and $n_d$, the task is to find $C_{T_i}, \forall$ ~i, i=1, 2, 3, 4 by solving Eq.~\ref{force_omega}. It can be seen that Eq.~\ref{force_omega} is not rational (the yawing moment equation), therefore it is difficult to explicitly obtain the values of $C_{T_i}$. To overcome this challenge, an additional loop that computes the desired rate of change in blade pitch angles for the given $T_d$, $l_d$, $m_d$, and $n_d$ is used. The objective of control allocation location loop is to determine $U=[\dot{C}_{T_1}\quad \dot{C}_{T_2}\quad \dot{C}_{T_3}\quad \dot{C}_{T_4}]^T$ that drives $[T\quad l\quad m\quad n]$ to $[T_d\quad l_d\quad m_d\quad n_d]$. For achieving this, the following second order stable error dynamics is chosen to synthesize the virtual control:
\begin{equation}
\left[\begin{array}{c}
\ddot{p}\\
\ddot{q}\\
\ddot{r}
\end{array}	\right] = 
\left[\begin{array}{c}
\ddot{p}_d\\
\ddot{q}_d\\
\ddot{r}_d
\end{array}	\right] + 2\zeta_{CA}{\omega_{n_{CA}}}^T
\left[\begin{array}{c}
\dot{p}_d - \dot{p}		\\
\dot{q}_d - \dot{q}		\\
\dot{r}_d - \dot{r}
\end{array}	\right] +\omega_{n_{CA}}{\omega_{n_{CA}}}^T
\left[\begin{array}{c}
p_d - p\\
q_d - q\\
r_d - r
\end{array}	\right]
\label{error_dyn_atT_di}
\end{equation} 
As the error dynamics are chosen on body rates, the desired body angular accelerations are obtained from Eq.~\ref{desired_body_euler_acc}. The actual body angular acceleration is obtained from Eq.~\ref{lmn_equ_invert}. By solving the above error dynamics, we get moment rates as:
\begin{equation}
\begin{bmatrix}
\dot{l}\\\dot{m}\\\dot{n}
\end{bmatrix}
=
\begin{bmatrix}
I_{xx}\ddot{p}\\I_{yy}\ddot{q}\\I_{zz}\ddot{r}
\end{bmatrix}
-
\begin{bmatrix}
(I_{yy} - I_{xx})(q\dot{r}+r\dot{q})\\(I_{zz} - I_{xx})(p\dot{r}+r\dot{p})\\(I_{xx} - I_{yy})(p\dot{q}+q\dot{p})
\end{bmatrix}
\label{moments_rate}
\end{equation}\\
Next, a first order error dynamics is applied on thrust to calculate its rate. The error dynamics is as follows
\begin{equation}
\dot{T} = k_p(T_d-T)
\label{thrusT_rate}
\end{equation}
where $k_p > 0$ is some proportionality constant. Using Eqs.~\ref{moments_rate} and~\ref{thrusT_rate},  $ \dot C_{T_i}$ are computed as follows
\begin{equation}
\begin{bmatrix}
\dot{C}_{T_1}\\\dot{C}_{T_2}\\\dot{C}_{T_3}\\\dot{C}_{t_4}
\end{bmatrix}
\label{thrusT_momenT_rates}
= \begin{bmatrix}
Kflag & Kflag & Kflag & Kflag\\
Kl & -Kl & -Kl & Kl\\
-flagKl & -flagKl & flagKl & flagKl\\
-flag\frac{3KR}{2}\sqrt{\frac{|c_{T1}|}{2}} & flag\frac{3KR}{2}\sqrt{\frac{|c_{t2}|}{2}} & -flag\frac{3KR}{2}\sqrt{\frac{|c_{t3}|}{2}} & flag\frac{3KR}{2}\sqrt{\frac{|c_{t4}|}{2}}
\end{bmatrix}^{-1} \begin{bmatrix}
\dot{T}\\\dot{l}\\\dot{m}\\\dot{n}
\end{bmatrix}
\end{equation}
Once virtual control input $U = [\dot{C}_{T_1}\quad \dot{C}_{T_2}\quad \dot{C}_{T_3}\quad \dot{C}_{T_4}]^T$ is obtained, it is integrated with the system dynamics to obtain the thrust coefficients. For the given thrust coefficients, the desired pitch angle can be obtained by solving Eq.~\ref{collective}. The control architecture block describing the control flow is shown in Fig.~\ref{control-scheme}.

\section{Numerical Results} \label{sec:NumericalResults}

In this section, the performance of the controller for a variable-pitch quadrotor is demonstrated through four examples. First, the stability of the inner loop is shown by stabilizing the perturbations given in the attitude from the hover state. Next, the overall performance is demonstrated by tracking a given trajectory. Next, the controller performs a flip maneuver on the quadrotor to enable inverted flight while maintaining altitude. Finally, the quadrotor tracks a sinusoidal trajectory in inverted state. The parameters used for numerical simulation of the variable pitch quadrotor are given in Table~\ref{quaddata}. Table~\ref{control_parameters} lists the control design parameters used for outer, inner and control allocation loops.

\begin{table} [h]
	\centering
	\caption{Parameters for variable pitch quadrotor used for numerical results}
	\begin{tabular}{ l  c  }
		\hline
		Mass of quadrotor, $M$	& 1.34 kg\\
		Radius of rotor blades, $R$	& 0.18 m\\
		Chord of rotor blades, $c$	& 0.03 m\\
		Distance of rotor axis from cg, $d$	& 0.3 m\\
		Airfoil lift curve slope, $C_{l_\alpha}$		& 5.23\\
		Airfoil drag coefficient, $C_{d_0}$		& 0.01\\
		Number of blades, $N_b$	&  2\\
		Rotational speed, $\Omega$&  282.7 rad/sec\\
		Moment of Inertia, $I_{xx}$& $1\times 10^{-3}$ kg-m/sec$^2$\\
		Moment of Inertia, $I_{yy}$& $1\times 10^{-3}$ kg-m/sec$^2$\\
		\hline
	\end{tabular}
	\label{quaddata}
\end{table}

\begin{table} [h]
	\centering
	\caption{Parameters used for control design}
	\begin{tabular}{ l  c  }
		\hline
		$\zeta_{out}$	& $[0.95 \quad 0.95 \quad 0.95]^T$\\
		$\omega_{n_{out}}$	& $[4.7 \quad 4.7 \quad 4.7]^T$\\
		$\zeta_{in}$	& $[0.92 \quad 0.92 \quad 0.92]^T$\\
		$\omega_{n_{in}}$	& $[30.5 \quad 30.5 \quad 20.5]^T$\\
		$\zeta_{CA}$	& $[0.91 \quad 0.91 \quad 0.91]^T$\\
		$\omega_{n_{CA}}$	& $[50 \quad 50 \quad 25]^T$\\
		$k_p$ & $10$\\
		
		\hline
	\end{tabular}
	\label{control_parameters}
\end{table}

\subsection{Attitude Stabilization}

\begin{figure}
	\centering
	\subfigure[Roll attitude]{
		\label{fig:ra}
		\includegraphics[width=0.37\textwidth]{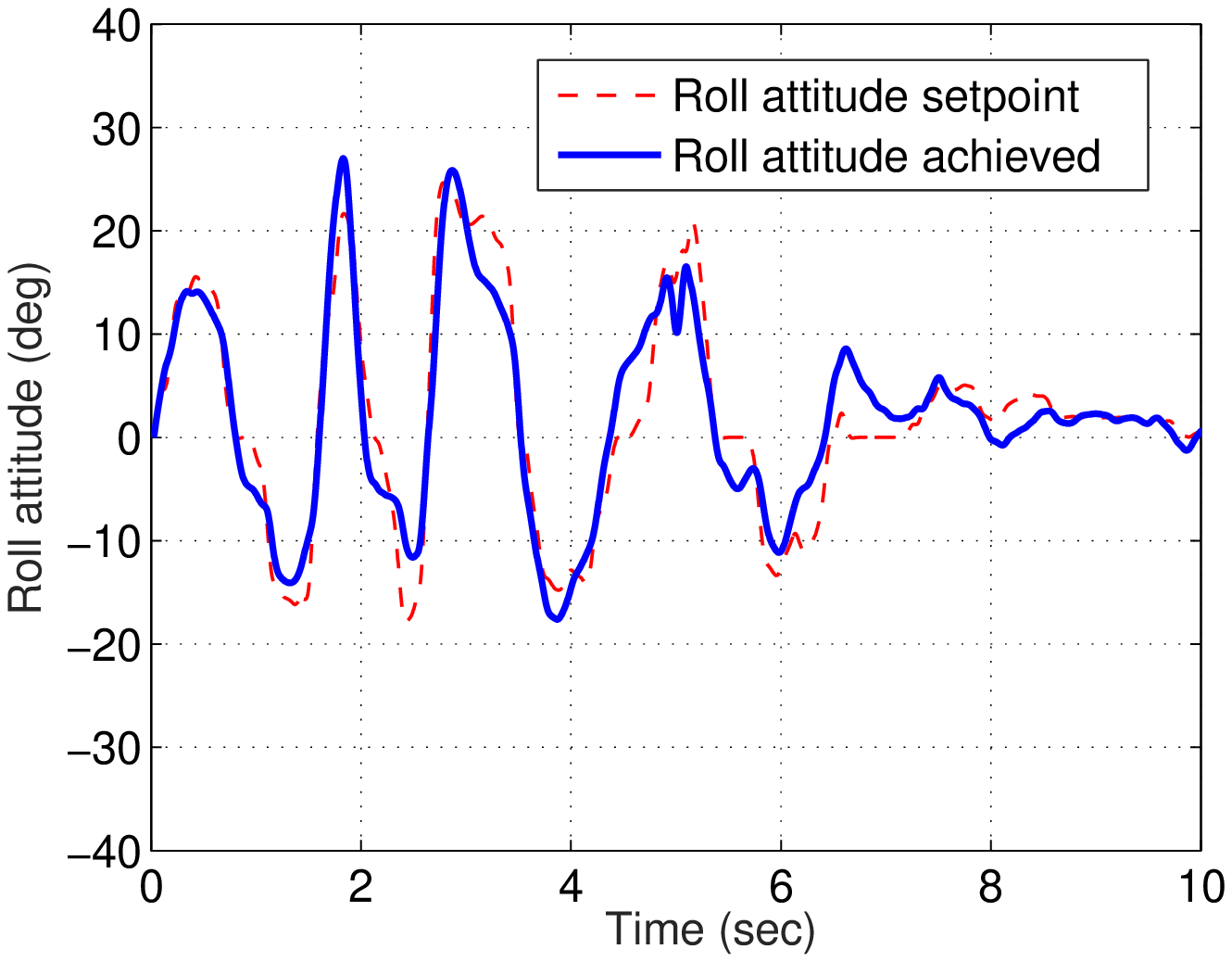}}
	\subfigure[Pitch attitude]{
		\label{fig:pa}
		\includegraphics[width=0.37\textwidth]{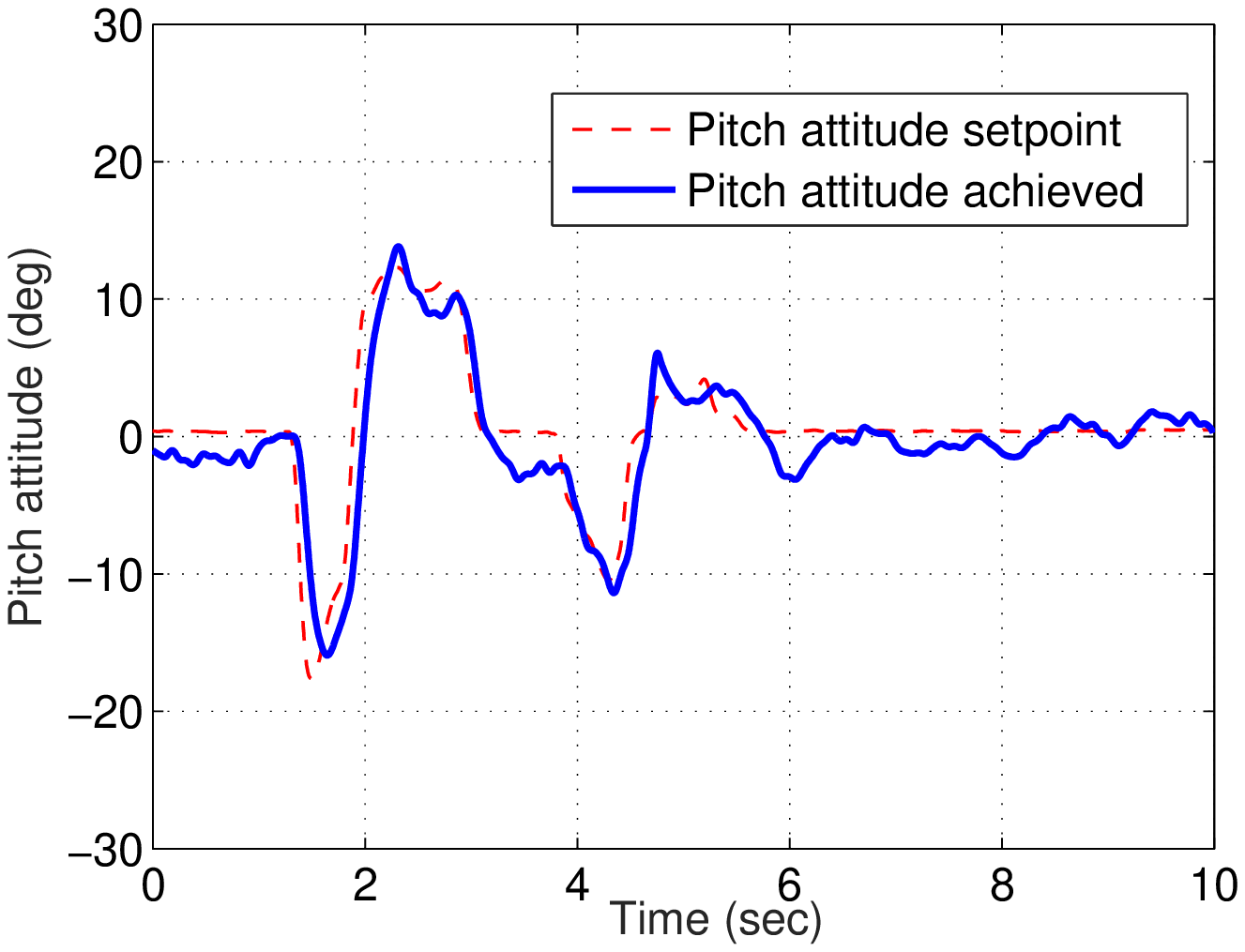}}
	\subfigure[Yaw attitude]{
		\label{fig:ya}
		\includegraphics[width=0.37\textwidth]{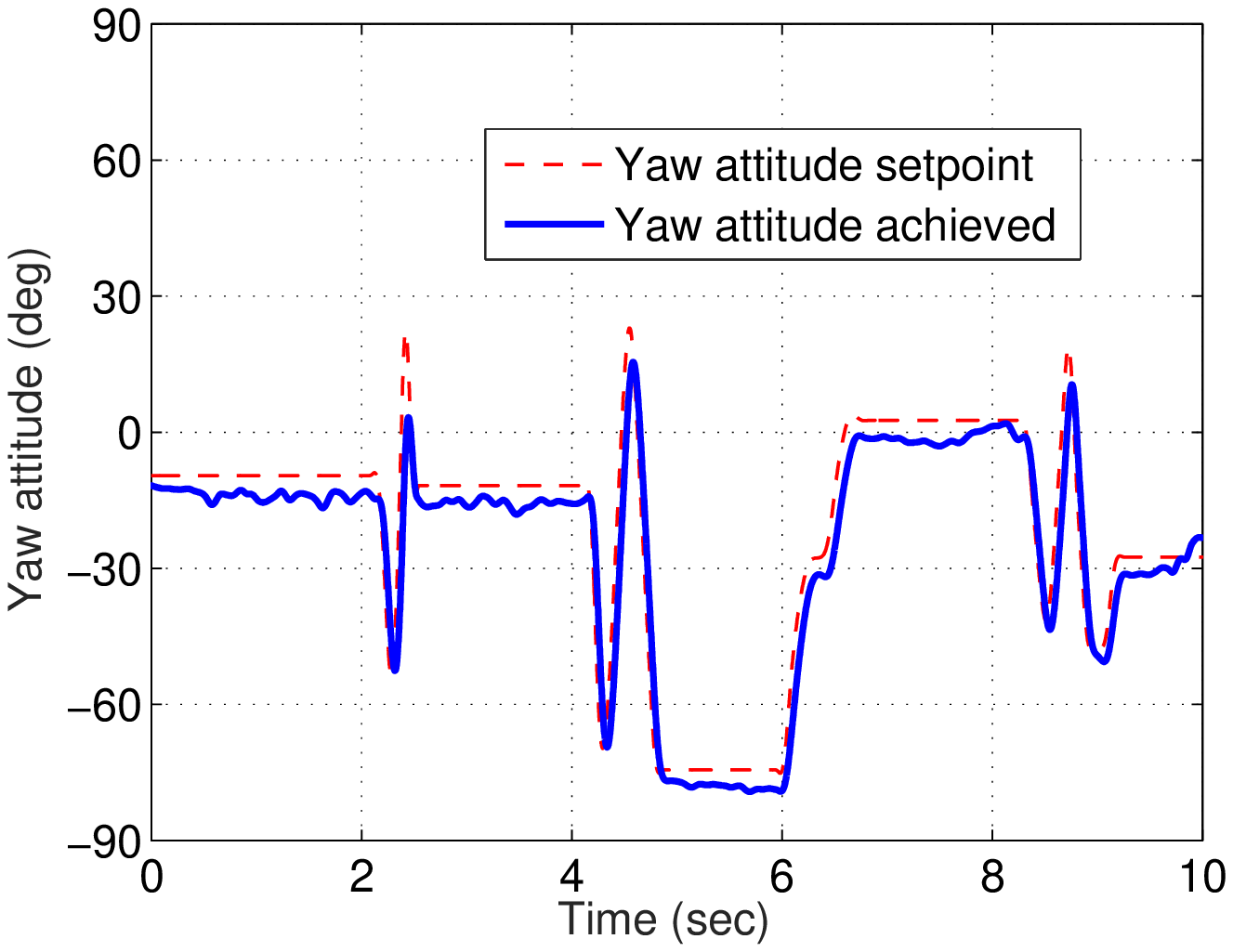}}
		\label{attitudevar}
\end{figure}

The variable-pitch quadrotor is a fairly new concept and therefore it is necessary to validate that vehicle attitude can be stabilized  changing blade pitch angle before we give simulation results. Toward this, a proof-of-concept single power plant electric powered variable pitch quadrotor UAV is designed. A PID controller based autopilot is developed and implemented on open source Pixhawk autopilot board to demonstrate attitude stabilization. The attitude controller  designed on inner loop  generates the desired roll, pitch, and yaw moments. The desired thrust is computed from the altitude stabilization. For control allocation, we assume that thrust and the moments are linear functions of blade pitch angle. This assumption makes $C_{T_i}$ calculation simple and enables the computation of the desired blade pitch angle using (\ref{modified_coeff_of_thrust}).  Figure~\ref{attitudevar} shows the attitude tracking performance for roll (Fig.~\ref{fig:ra}), pitch (Fig.~\ref{fig:pa}) and yaw (Fig.~\ref{fig:ya}) attitudes  during closed-loop flight test of the proof-of-concept UAV. It can be observed that the controller is able to accurately track the commanded setpoints for each of the vehicle attitudes. The setpoints during the flight test are being provided by human pilot through a joy-stick.

\begin{figure}
	\centering
	\subfigure[Attitude]{
		\label{attitude_stabilization}
		\includegraphics[width = 0.45\textwidth]{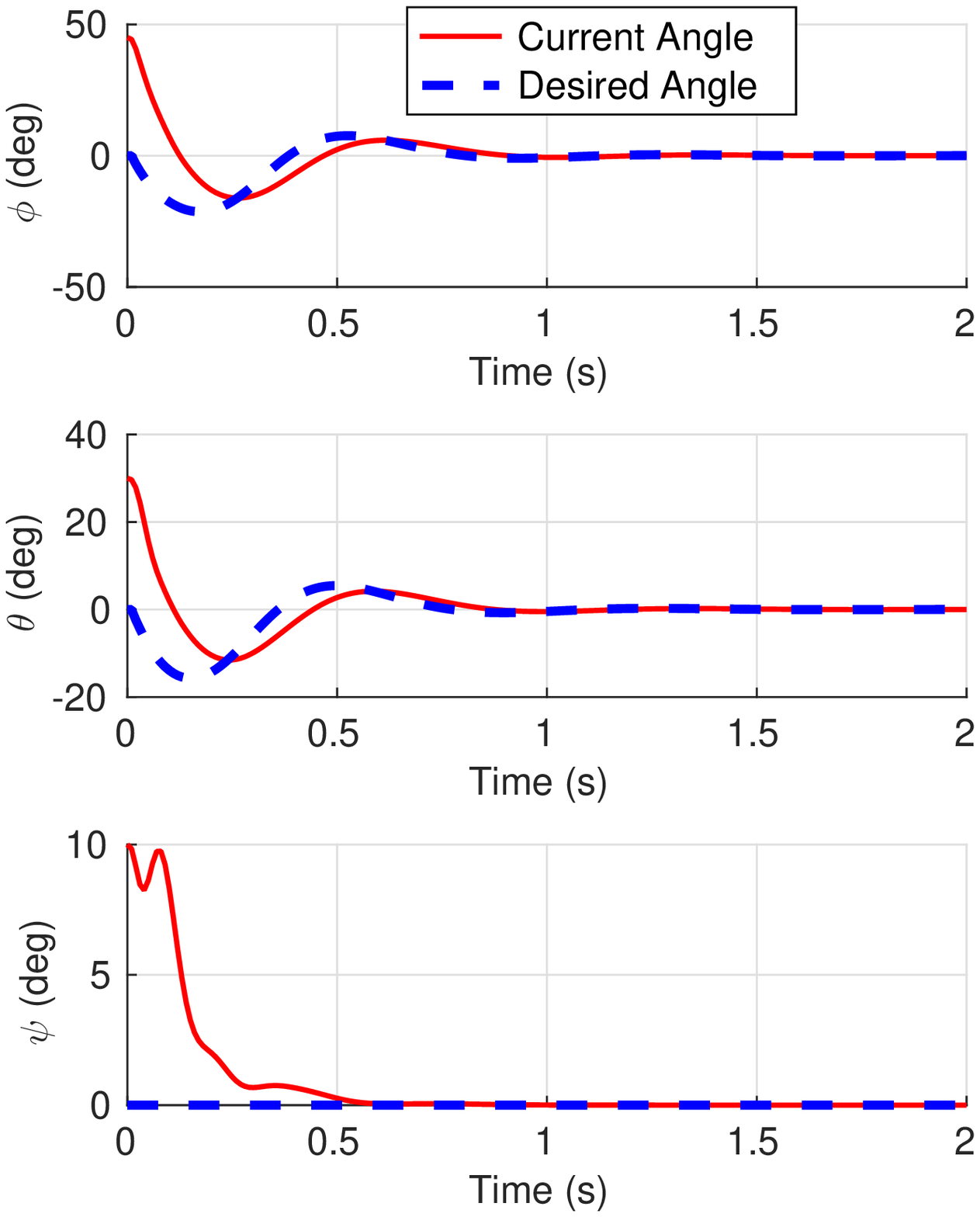}} 
	\centering
	\subfigure[Position]{
		\label{position_stabilization}
		\includegraphics[width = 0.45\textwidth]{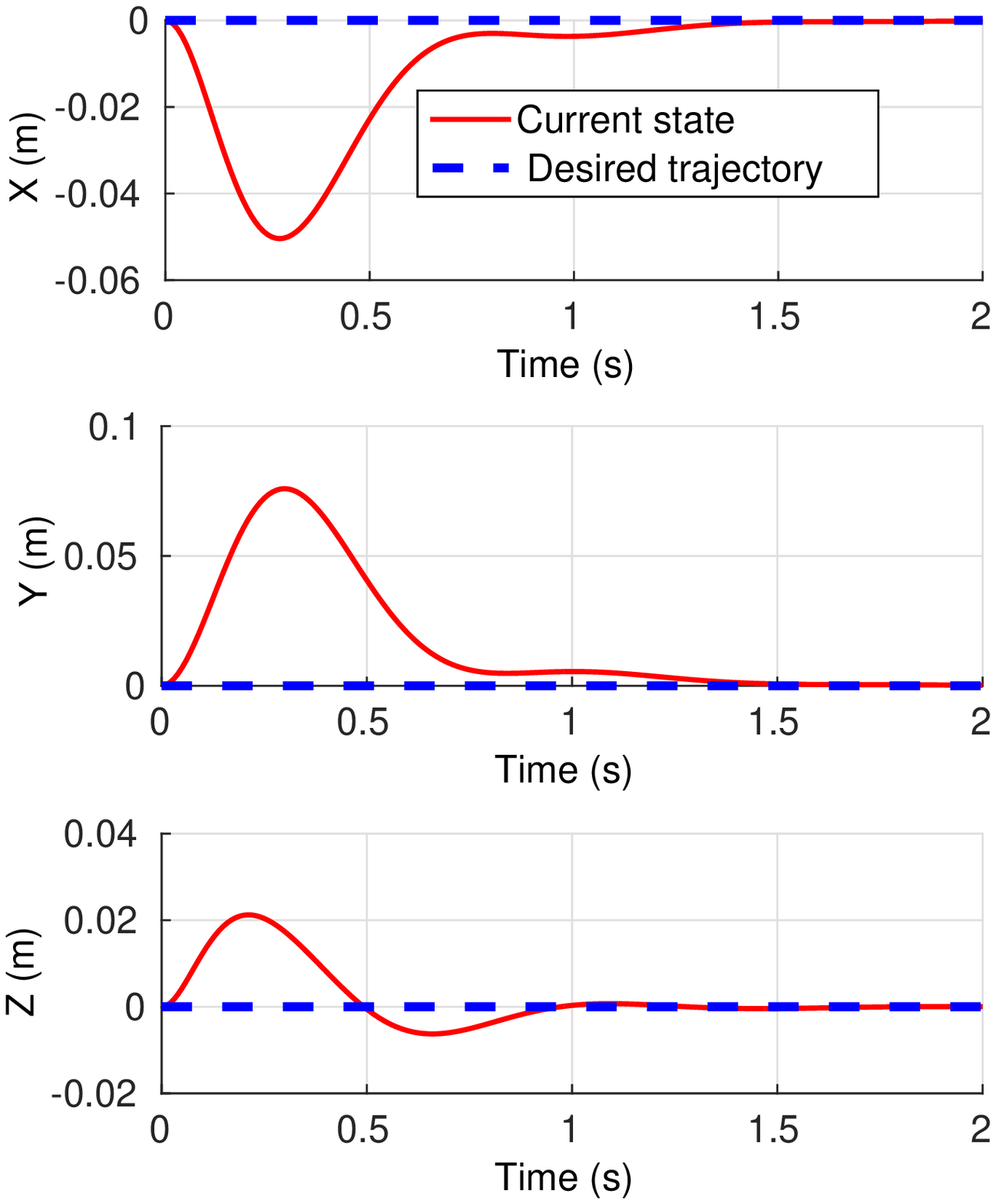}} 
	\caption{\textbf{Position and attitude variation during quadrotor stabilization}}
	\label{fig:stabilization}
\end{figure}

\begin{figure}
	\centering
	\includegraphics[width = 0.5\textwidth]{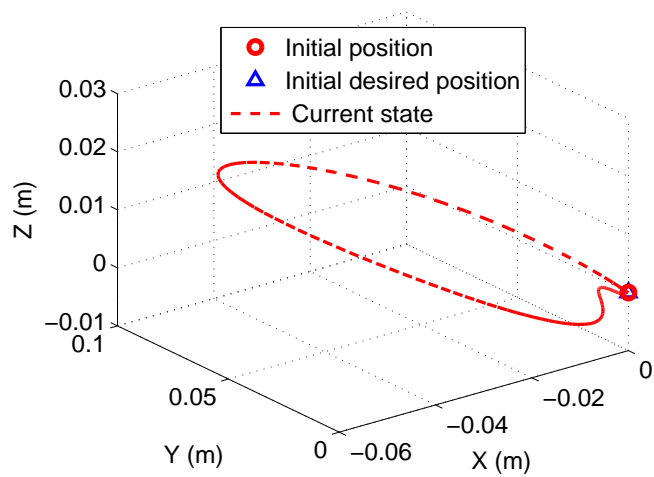}
	\caption{\textbf{Variation in position shown in three-dimensions during quadrotor stabilization}}
	\label{3d_stabilization}
\end{figure}

\subsection{Position Stabilization}

\begin{figure}
	\centering
	\includegraphics[width = 0.9\textwidth]{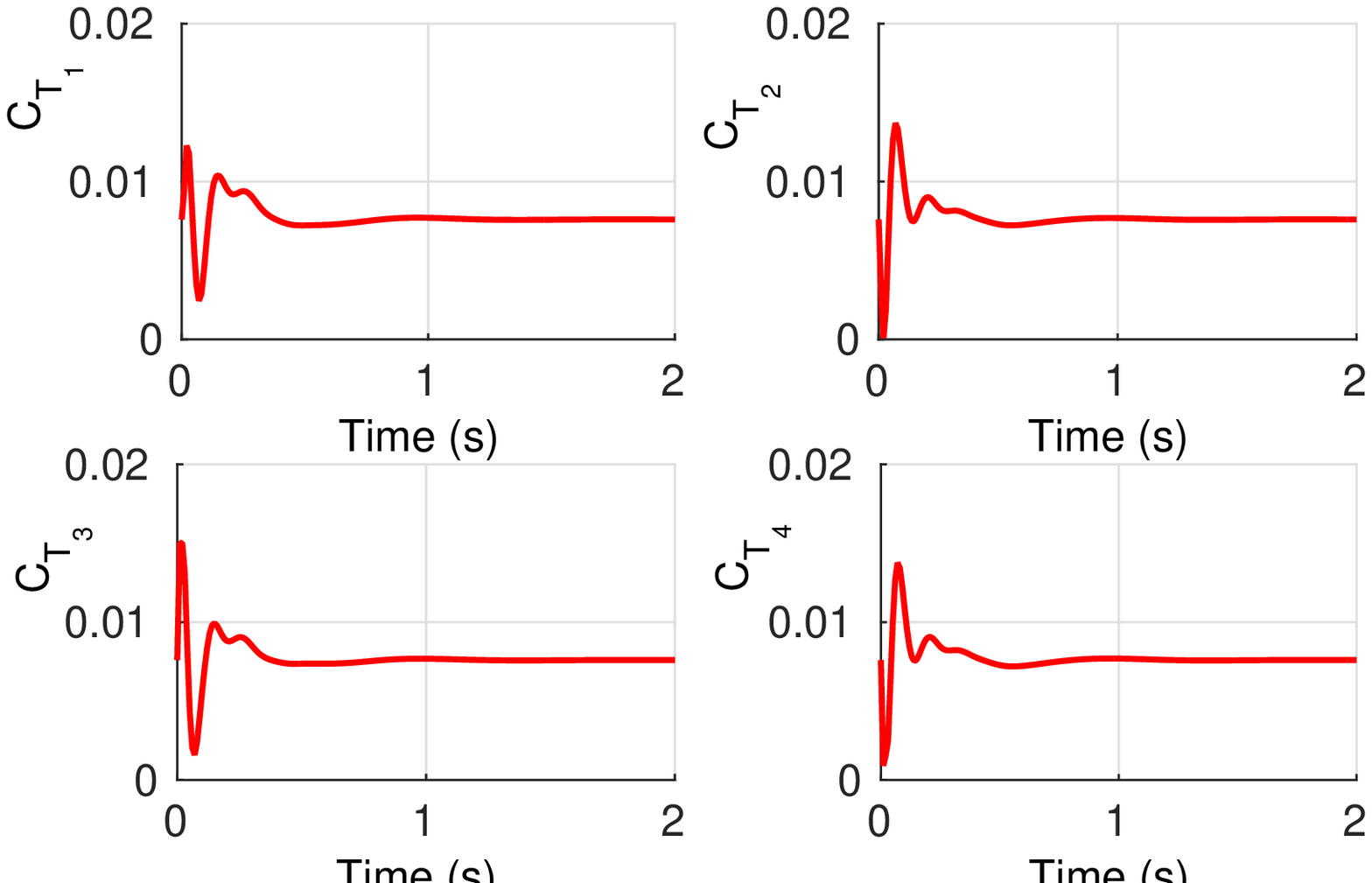}
	\caption{\textbf{Time history of required thrust coefficients for stabilizing the quadrotor}}
	\label{coeff_of_thrust_stabilization}
\end{figure}

\begin{figure}
	\centering
	\includegraphics[width = 0.9\textwidth]{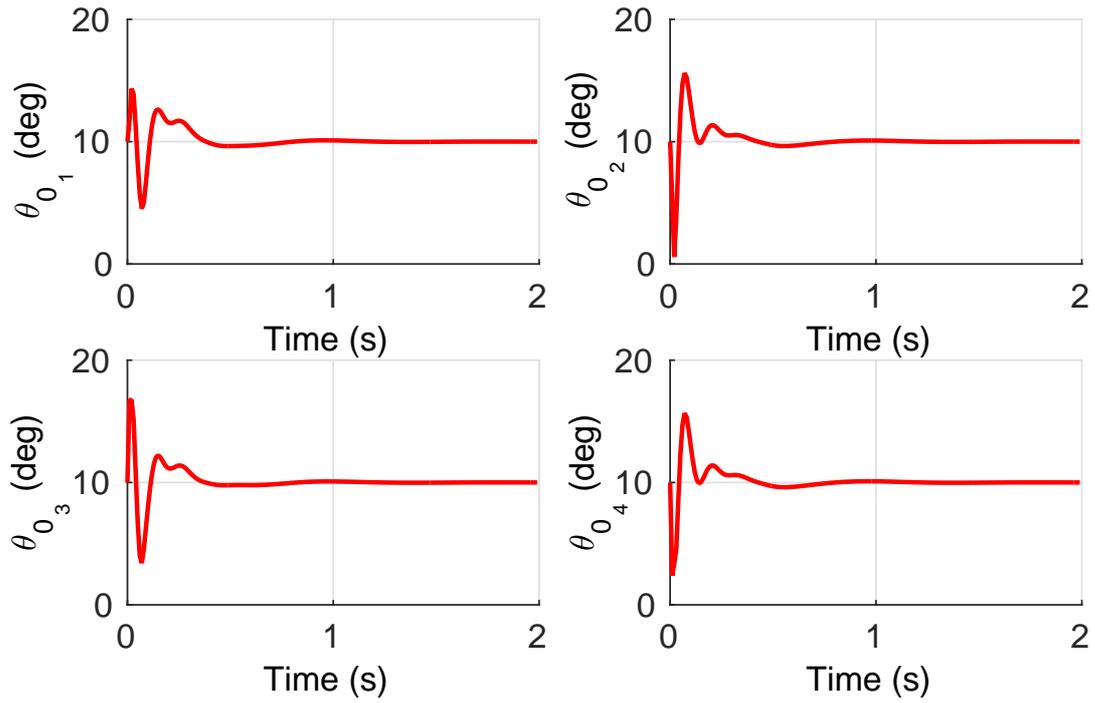}
	\caption{\textbf{Time history of required collective pitch inputs for stabilizing the quadrotor}}
	\label{collective_stabilization}
\end{figure}

To demonstrate the ability of the controller to stabilize and maintain vehicle position in the event of disturbance, the initial values of roll ($\phi$), pitch ($\theta$), and yaw ($\psi$) angles are perturbed by $45^\circ$, $30^\circ$ and $10^\circ$, respectively. The controller brings back the vehicle to hover attitude by reducing the given perturbations quickly. The time history of attitude variation during this process is shown in Fig.~\ref{attitude_stabilization}. The act of stabilizing the quadrotor against large disturbance results in rapid changes in individual rotor thrust resulting in change in position of the quadrotor. But, once the attitude disturbance is controlled back to the desired state, the deviation in position is also reduced to zero. The time history of variation in $x$, $y$, and $z$ position coordinates during the stabilization is shown in Fig.~\ref{position_stabilization}. It is observed that the attitude is stabilized in less than one sec and the position is restored in less than 1.5 sec. The overall variation in position in three-dimensions is shown in Fig.~\ref{3d_stabilization} and is observed to be small. The time history of variation of coefficient of thrust $C_T$ for individual rotors required for stabilizing the quadrotor is shown in Fig.~\ref{coeff_of_thrust_stabilization}. The corresponding collective pitch input required to achieve this thrust coefficient is shown in Fig.~\ref{collective_stabilization}. Even though the vehicle is released from an attitude which is significantly disturbed from the desired hover attitude, the stabilization of the quadrotor is achieved with moderate control actuation (less than $16^\circ$). Next, the performance of the controller in trajectory tracking is demonstrated.

\subsection{Trajectory Tracking}
\begin{figure}
	\centering
	\subfigure[Attitude]{
		\label{attitude_trajectory}
		\includegraphics[width = 0.45\textwidth]{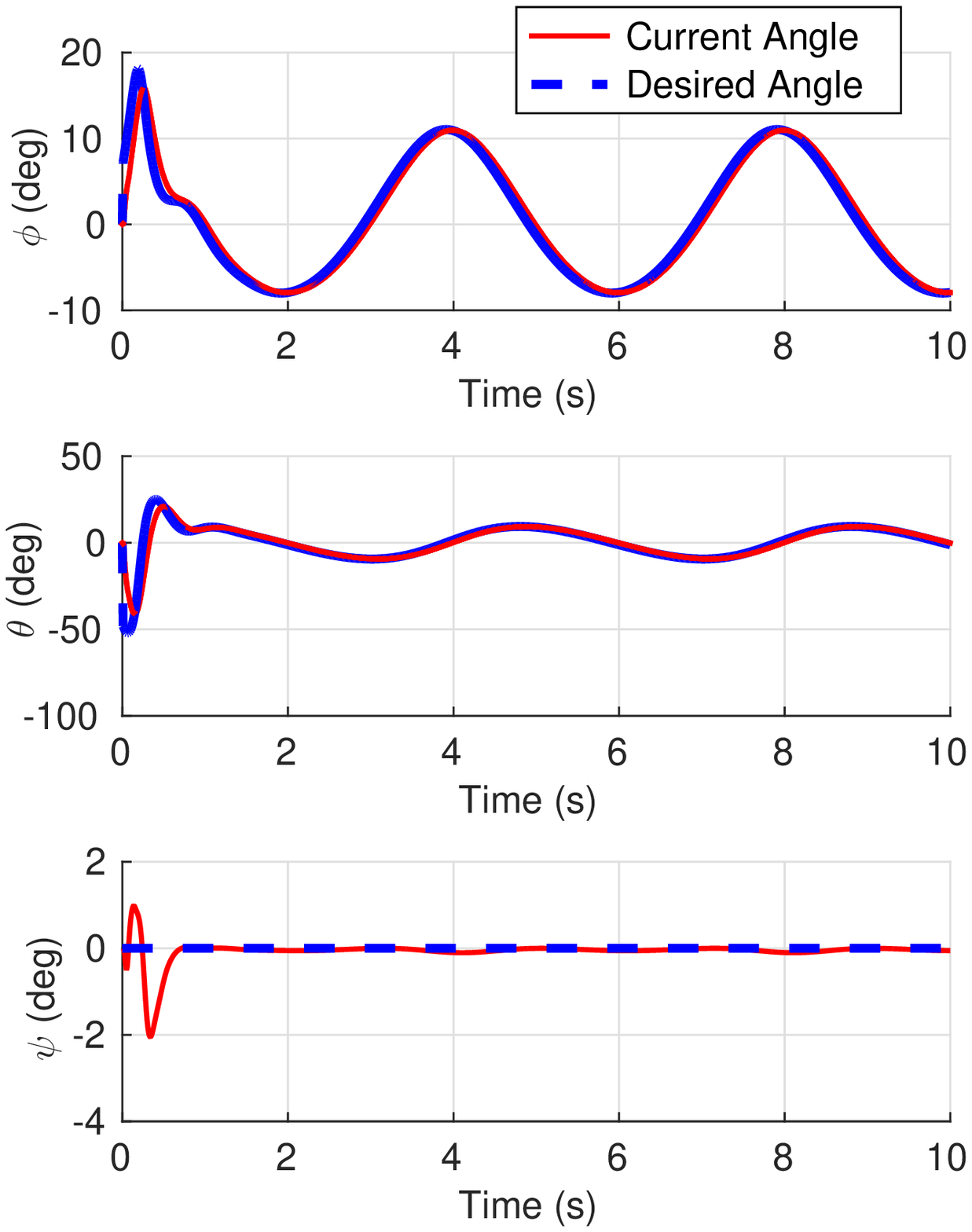}} 
	\centering
	\subfigure[Position]{
		\label{position_trajectory}
		\includegraphics[width = 0.45\textwidth]{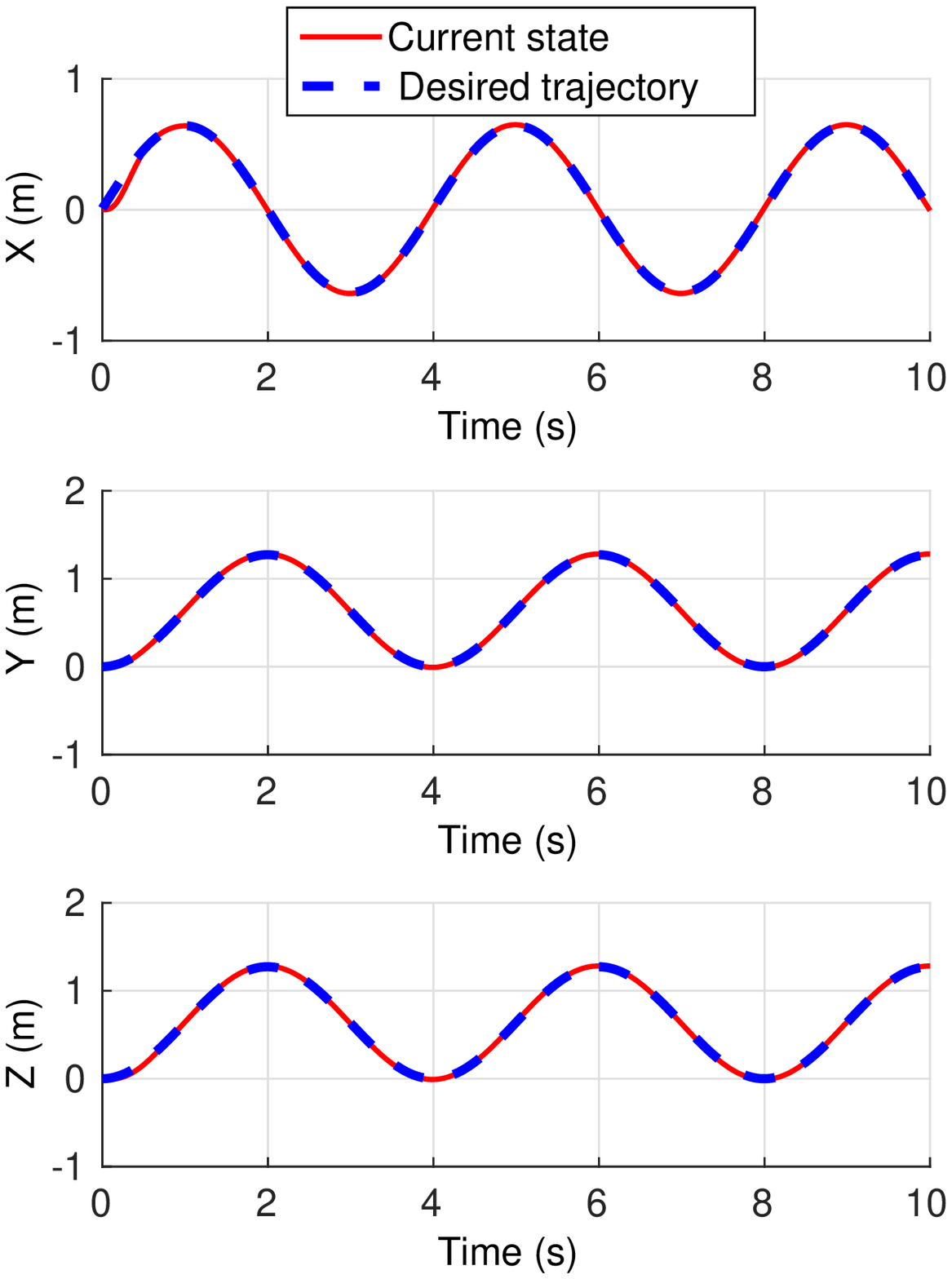}} 
	\caption{\textbf{Position and attitude variation during tracking of prescribed trajectory by quadrotor}}
	\label{fig:trajectory}
\end{figure}

The trajectory tracking capability of the controller is evaluated by commanding it to follow a sinusoidal path. The initial position of the quadrotor is set as the origin (0,0,0),  the attitude angles ($\phi$, $\theta$, $\psi$)  are  $(0^{\circ},0^{\circ},0^{\circ})$. The quadrotor is commanded from this position to follow a sinusoidal input of $sin(\frac{\pi}{2}t)$ meters in $X$, $Y$, and $Z$ directions. The controller is able to track the trajectory accurately as observed from the time history of variation of attitude and position during the trajectory tracking shown in Figs.~\ref{attitude_trajectory} and~\ref{position_trajectory}. In order to follow the given trajectory, the outer loop generates required roll ($\phi$) and pitch ($\theta$) commands which are shown by dashed line in Fig.~\ref{attitude_trajectory}. The red line shows the tracking of desired command.

\begin{figure}
	\centering
	\includegraphics[width = 0.5\textwidth]{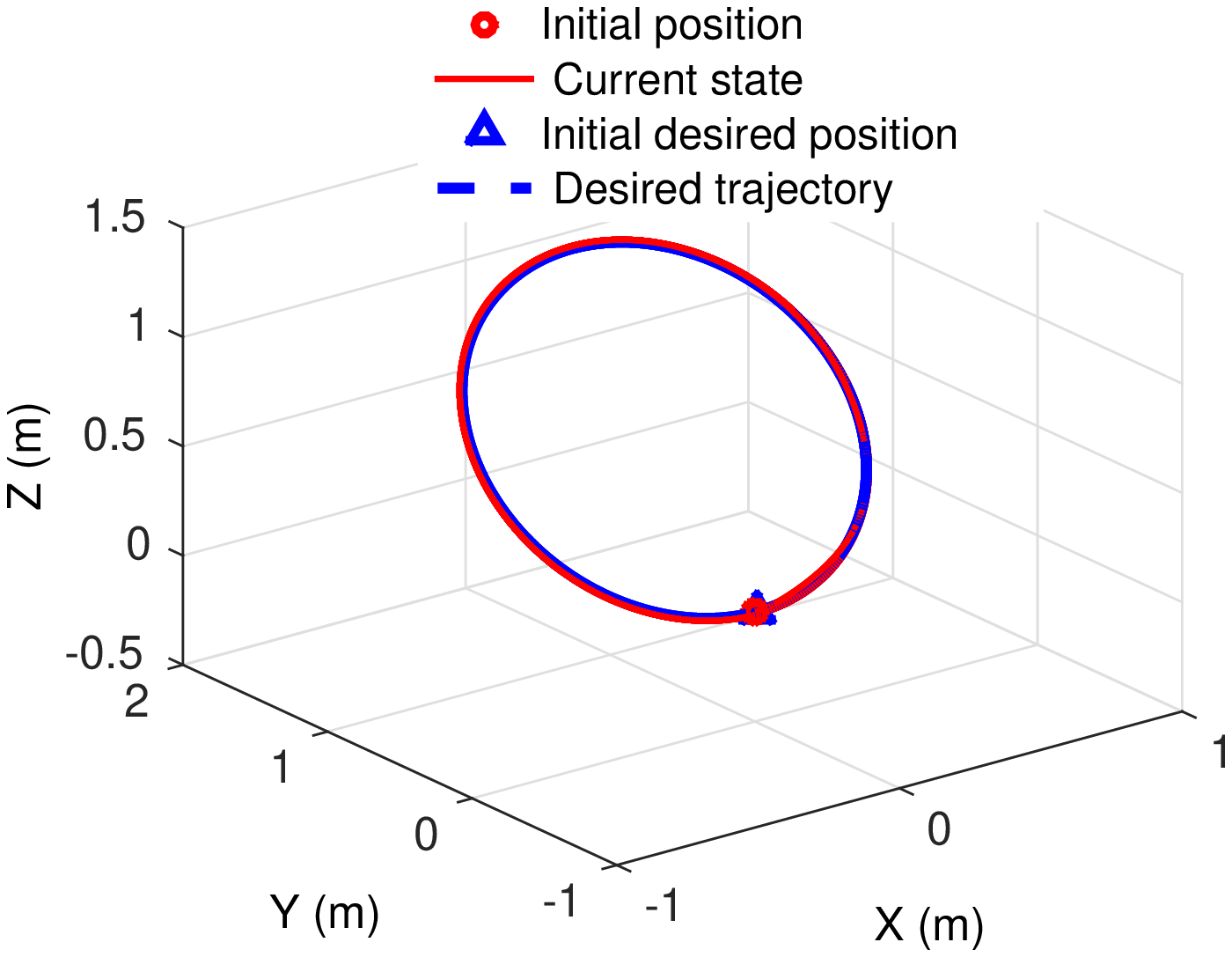}
	\caption{\textbf{Variation in position of quadrotor shown in three-dimensions during tracking of prescribed trajectory}}
	\label{3d_trajectory}
\end{figure}

\begin{figure}
	\centering
	\includegraphics[width = 0.9\textwidth]{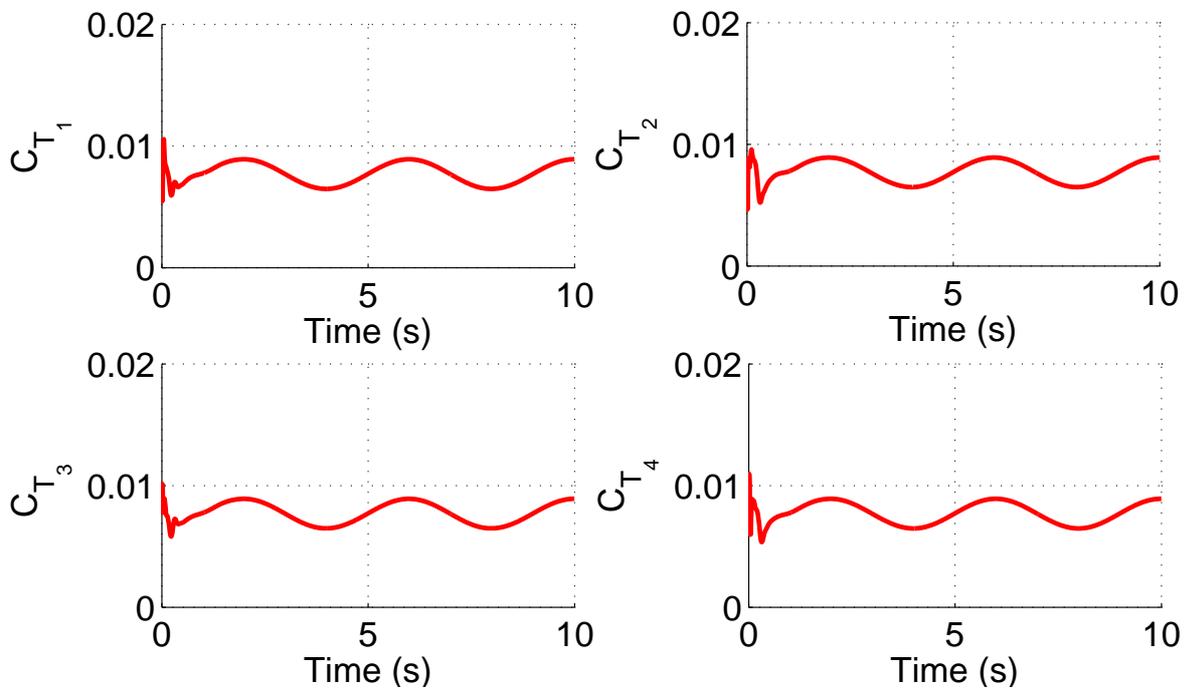}
	\caption{\textbf{Time history of required thrust coefficients for tracking prescribed trajectory}}
	\label{coeff_of_thrust_trajectory}
\end{figure}

\begin{figure}
	\centering
	\includegraphics[width = 0.9\textwidth]{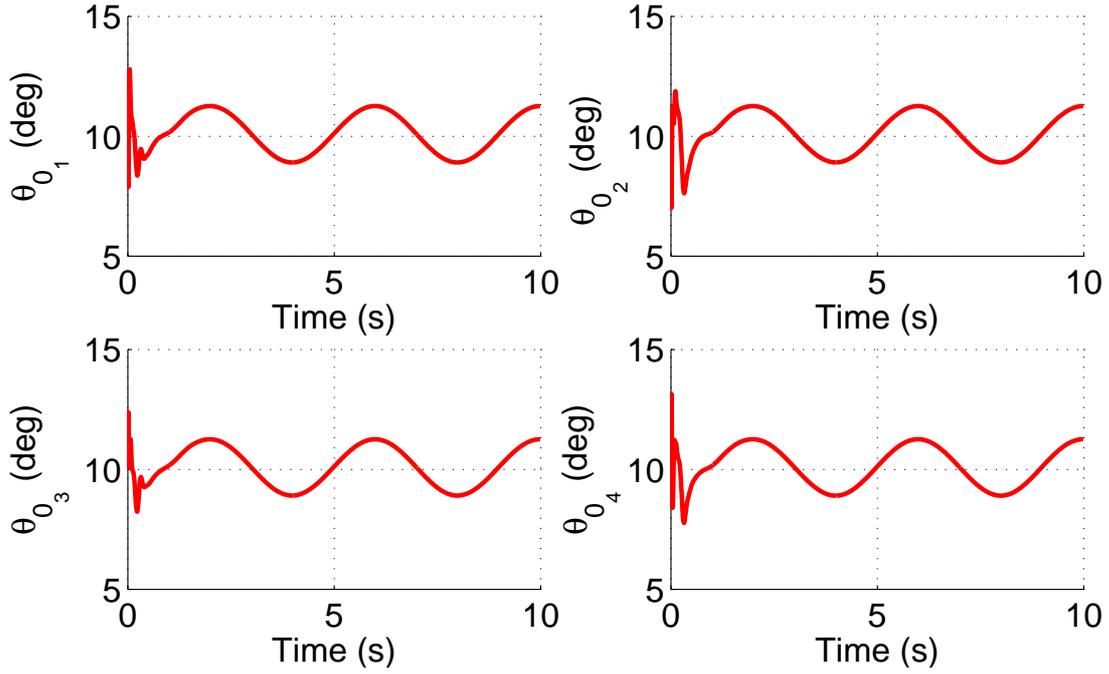}
	\caption{\textbf{Time history of required collective pitch inputs for tracking prescribed trajectory}}
	\label{collective_trajectory}
\end{figure}

The trajectory traced by the quadrotor in three-dimensions is shown in Fig.~\ref{3d_trajectory}, which is a slanted circle. The given command is shown by dashed line and the accurately tracked trajectory is shown using solid lines. Since, the $x$, $y$, and $z$ coordinates depicting the location of the vehicle are varying sinusoidally, it is expected that the controller input would also vary sinusoidally as shown in Figs.~\ref{coeff_of_thrust_trajectory} and~\ref{collective_trajectory}. Again, it can observed from these figures that the controller is able to regulate the blade pitch angles and thereby generate required thrust to track the prescribed trajectory.

\begin{figure}
	\centering
	\subfigure[Attitude]{
		\label{attitude_flip}
		\includegraphics[width = 0.45\textwidth]{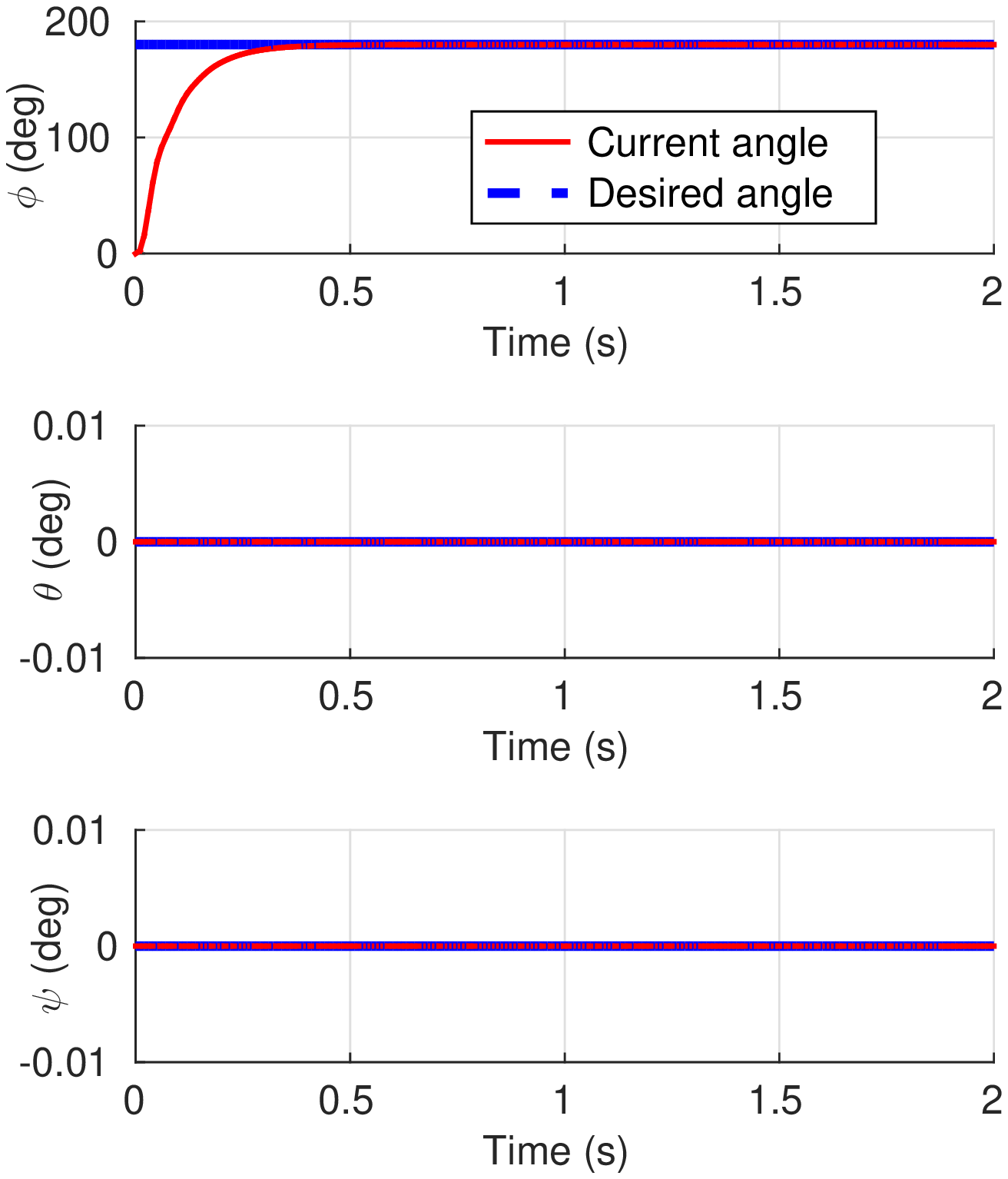}} 
	\centering
	\subfigure[Position]{
		\label{position_flip}
		\includegraphics[width = 0.45\textwidth]{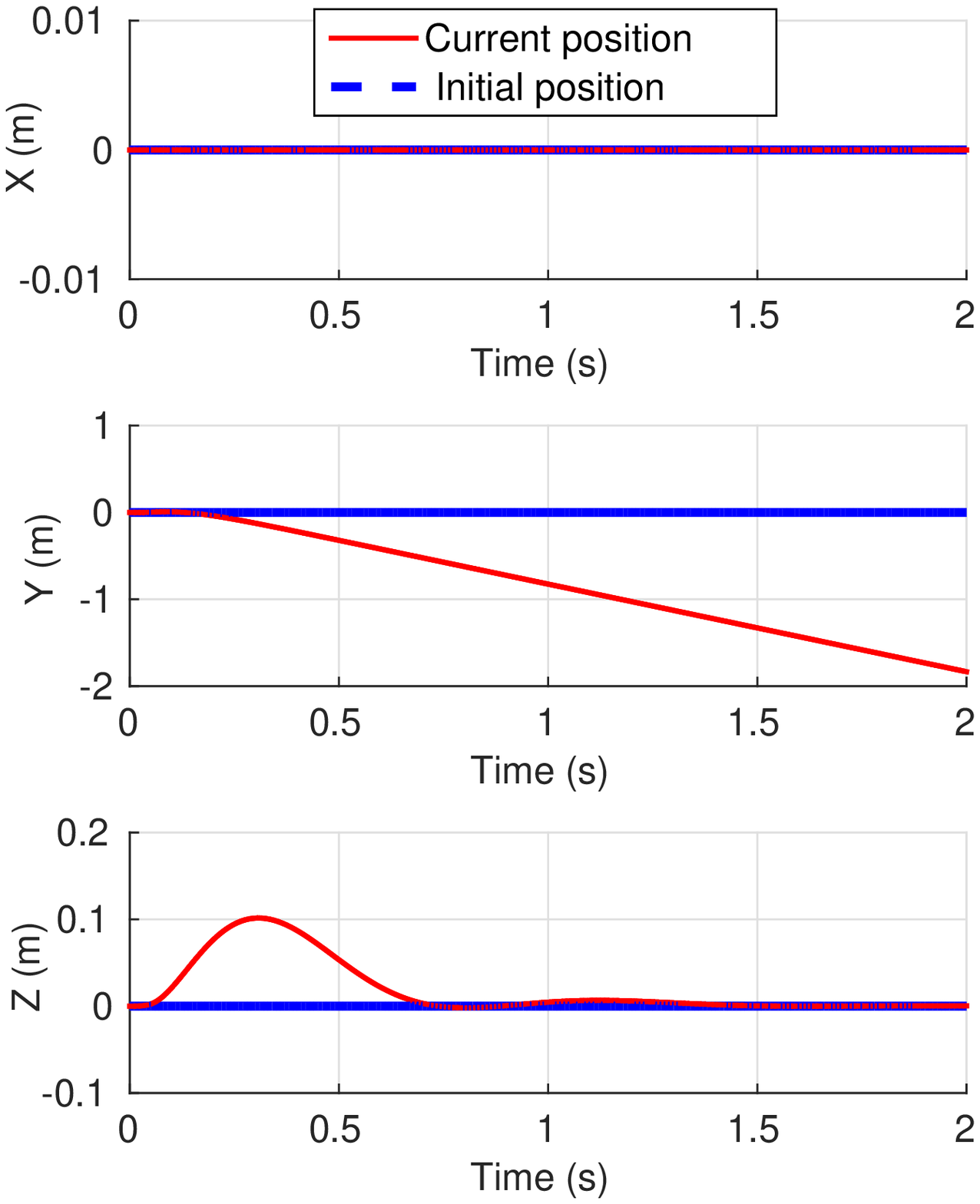}} 
	\caption{\textbf{Position and attitude variation during flip maneuver performed by quadrotor}}
	\label{fig:flip}
\end{figure}

\subsection{Flip Maneuver}

In this section, the capability of the variable pitch quadrotor and the developed controller is demonstrated by performing a complicated flip maneuver. In this maneuver, the quadrotor is simulated to fly upside down starting from the stable hover position with roll, pitch and yaw attitude angles  maintained at $0^{\circ}$. The controller then commands the quadrotor to change the roll angle to $180^{\circ}$ while maintaining the pitch and yaw angles. The time history of commanded and achieved attitude angles of the vehicle is shown in Fig.~\ref{attitude_flip}. The time history of position of the quadrotor during the flip maneuver is shown in Fig. ~\ref{position_flip}. During this maneuver no attempt is made to control the position of the quadrotor and only attitude is targeted. As a consequence, it can be observed that during the transition from $0^{\circ}$ roll attitude to $180^{\circ}$ roll attitude, the quadrotor generates some acceleration which results in a small velocity along lateral ($Y$) direction which makes the $Y$ coordinate position to increase with time. 

\begin{figure}
	\centering
	\includegraphics[width = 0.8\textwidth]{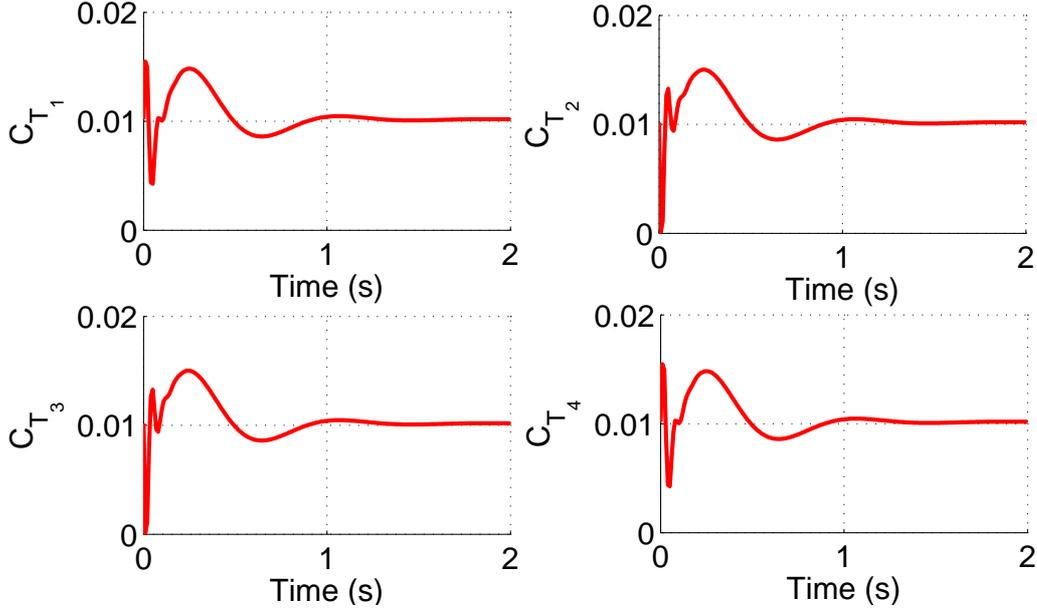}
	\caption{\textbf{Time history of required thrust coefficients for performing flip maneuver}}
	\label{coeff_of_thrust_flip}
\end{figure}

\begin{figure}
	\centering
	\includegraphics[width = 0.8\textwidth]{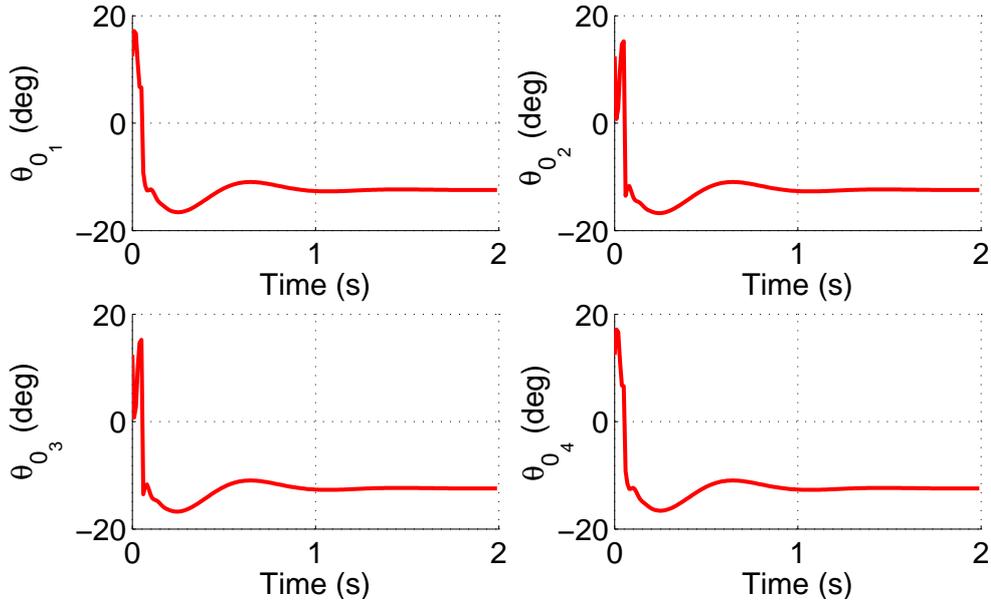}
	\caption{\textbf{Time history of required collective pitch input for performing flip maneuver}}
	\label{collective_flip}
\end{figure}

\begin{figure}
	\centering
	\includegraphics[width =\textwidth]{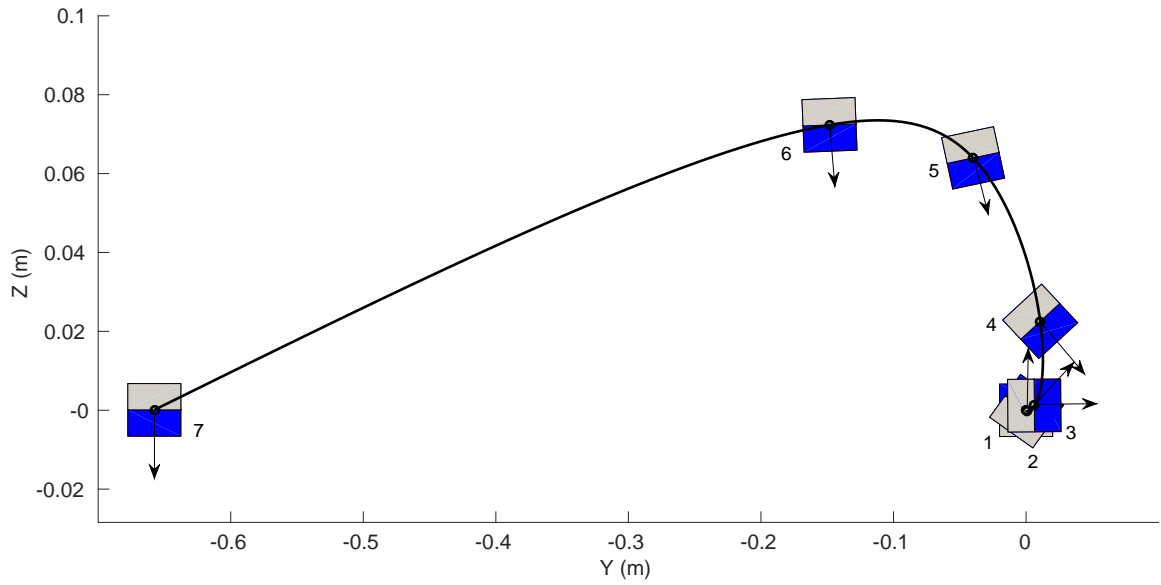}
	\caption{\textbf{Trajectory in Y-Z plane during flipping}}
	\label{flip_flip}
\end{figure}

After the execution of the flip maneuver, all the four rotors of the quadrotor produce thrust of equal magnitude and has same value as that of thrust in hover mode (see Fig.~\ref{coeff_of_thrust_flip}). As expected, the rotors of the quadrotor, when in inverted flight, operate at negative collective pitch angles to generate thrust in upward direction as shown in Fig.~\ref{collective_flip}. The trajectory of the quadrotor during the flip maneuver in $Y-Z$ plane is shown in Fig. ~\ref{flip_flip}. The upright attitude of the quadrotor at its original location is marked by number `1' and is depicted using a square with dark shade in the top half and light shade in the bottom half portion. The snapshots of the simulated flip maneuver are marked by numbers `1' through `7'. It is observed that the flipping of the quadrotor is completed by the time the quadrotor reaches location `6' as it attains upside down attitude marked by a square with bottom half in dark and top half in light shade. The centre of mass of the quadrotor is observed to move by only 0.14 m in lateral direction and 0.07 m in vertical direction during the execution of the flip maneuver. The quadrotor maintains its altitude but drifts in Y-direction due to the reason explained above. 

\begin{figure}
	\centering
	\subfigure[Attitude]{
		\label{attitude_inverted}
		\includegraphics[width = 0.45\textwidth]{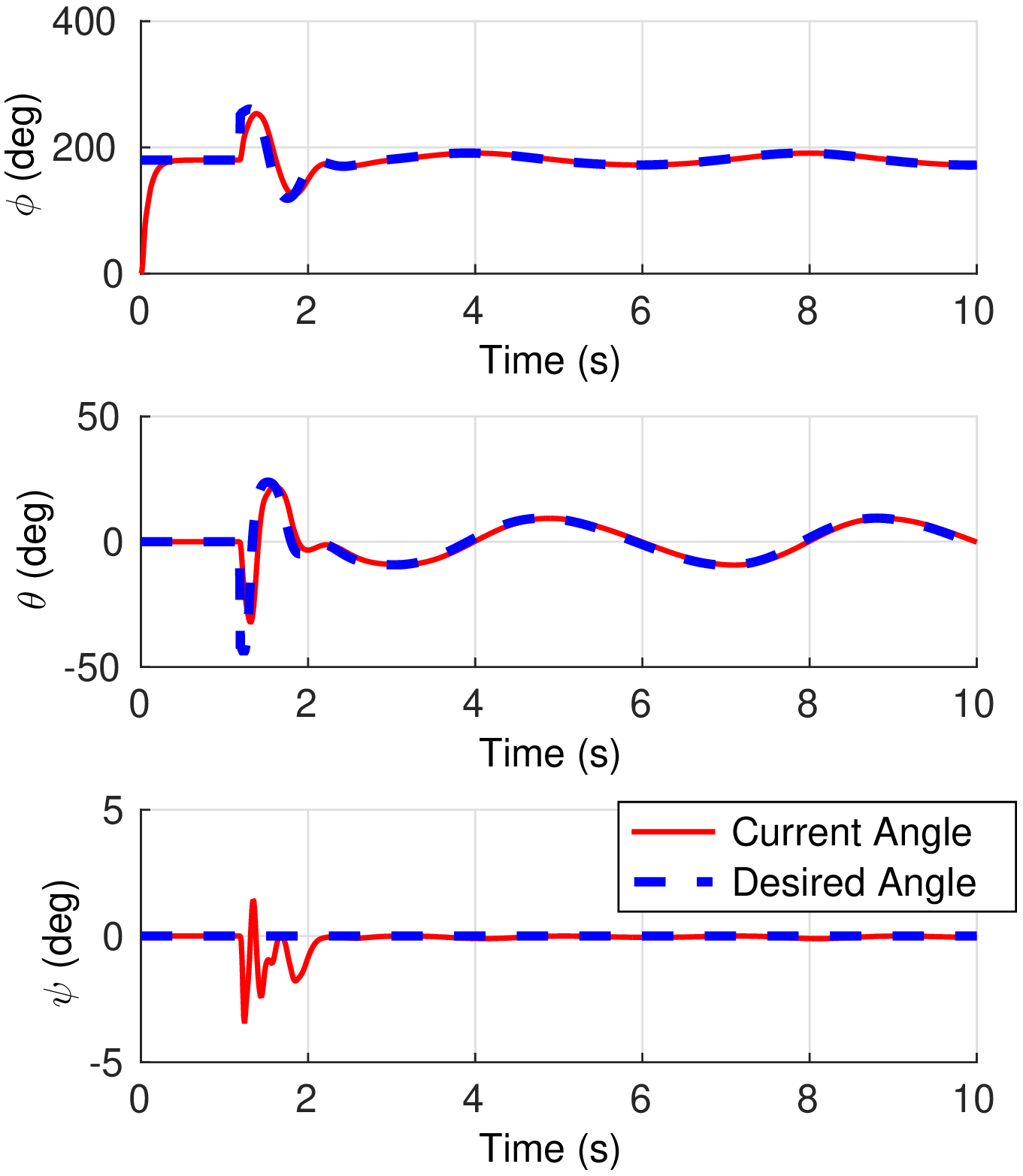}} 
	\centering
	\subfigure[Position]{
		\label{position_inverted}
		\includegraphics[width = 0.45\textwidth]{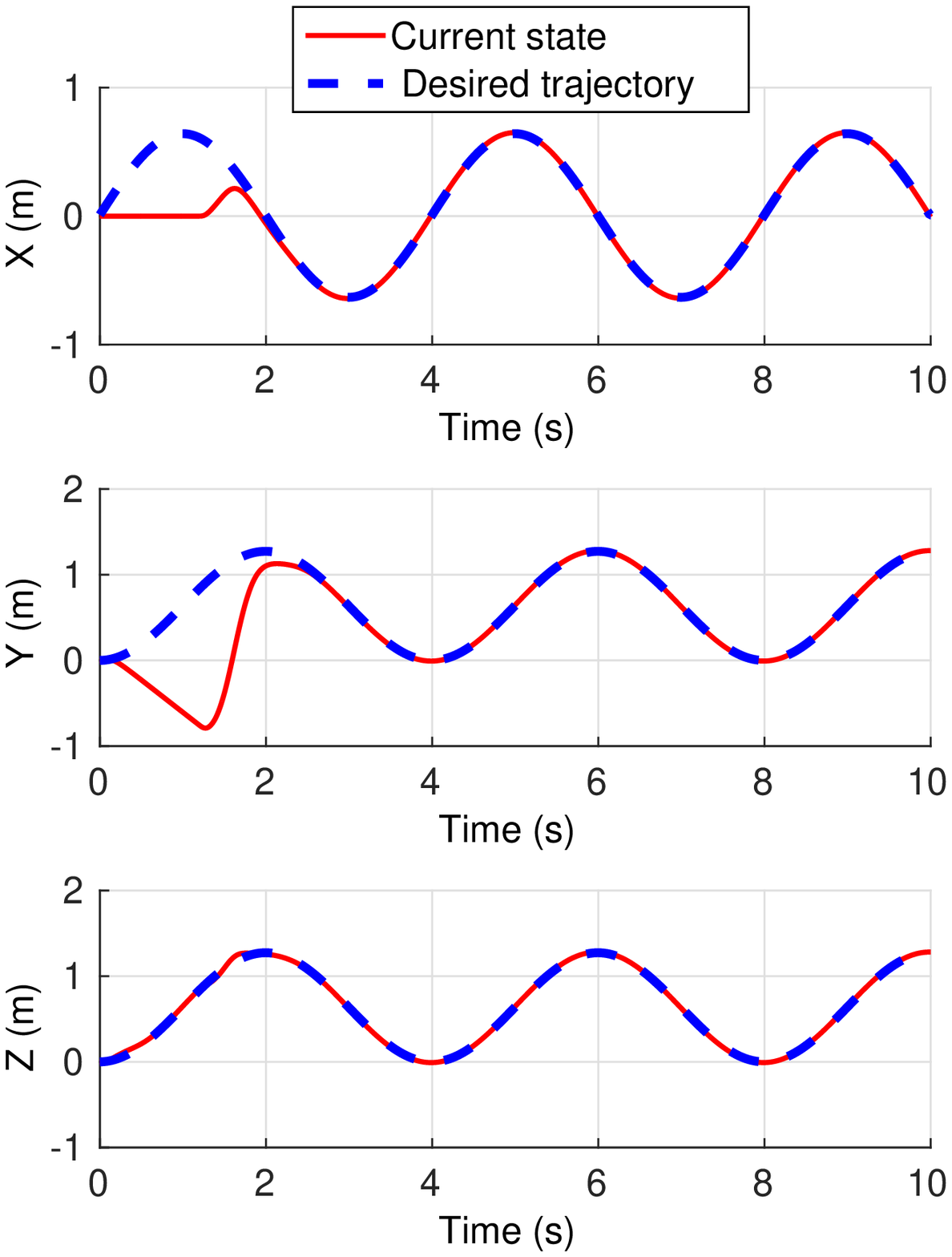}} 
	\caption{\textbf{Position and attitude variation during trajectory tracking with inverted quadrotor}}
	\label{fig:invert}
\end{figure}

\subsection{Inverted Flight}

For final demonstration of the performance of the controller, a sinusoidal trajectory is tracked by the quadrotor in inverted orientation. Starting with hover, the quadrotor is commanded to perform a flip maneuver followed by tracking of a sinusoidal trajectory of $sin(\frac{\pi}{2}t)$ meters in $X$, $Y$, and $Z$ directions. 

Similar to trajectory following, the attitude required to track the trajectory is shown by dashed line in Fig.~\ref{attitude_inverted} and actual attitude attained is shown using solid line. Fig.~\ref{attitude_inverted} shows that quadrotor flips within 1 sec attaining a roll angle of $180^\circ$, and then starts following the desired attitude to track the prescribed trajectory. The time history of desired and tracked positions are shown in Fig.~\ref{position_inverted}. After the initial deviation of $X$ and $Y$ location during the flipping motion, the inverted quadrotor is able to track the desired trajectory with great precision. The corresponding flight path in three-dimensions is shown in Fig.~\ref{3d_trajectory}. 

\begin{figure}[ht]
	\centering
	\includegraphics[width = 0.8\textwidth]{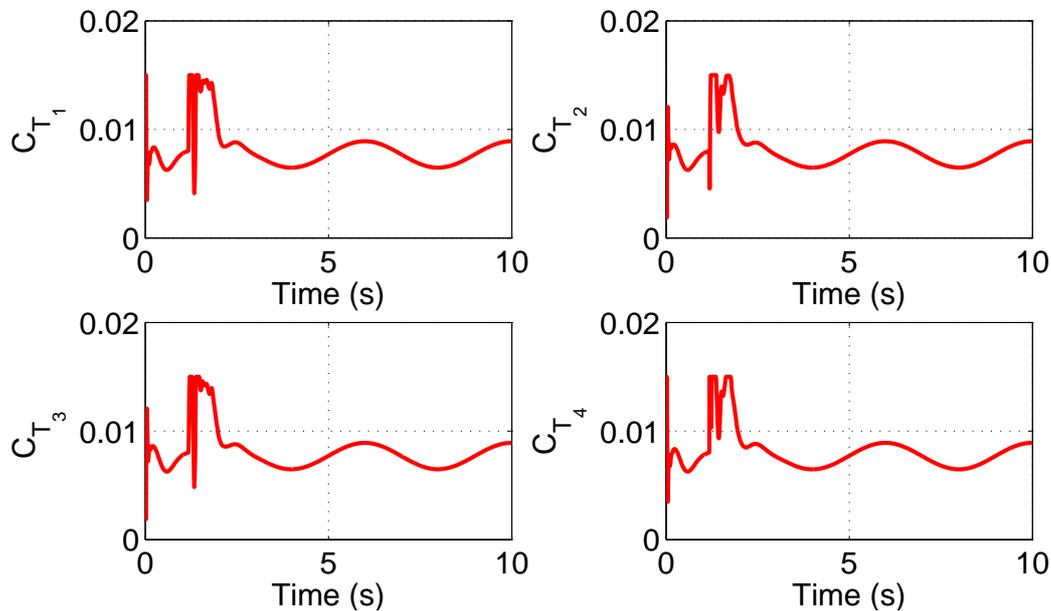}
	\caption{\textbf{Required values of thrust coefficients variation for tracking trajectory during inverted flight}}
	\label{coeff_of_thrust_inverted}
\end{figure}

\begin{figure}[ht]
	\centering
	\includegraphics[width = 0.8\textwidth]{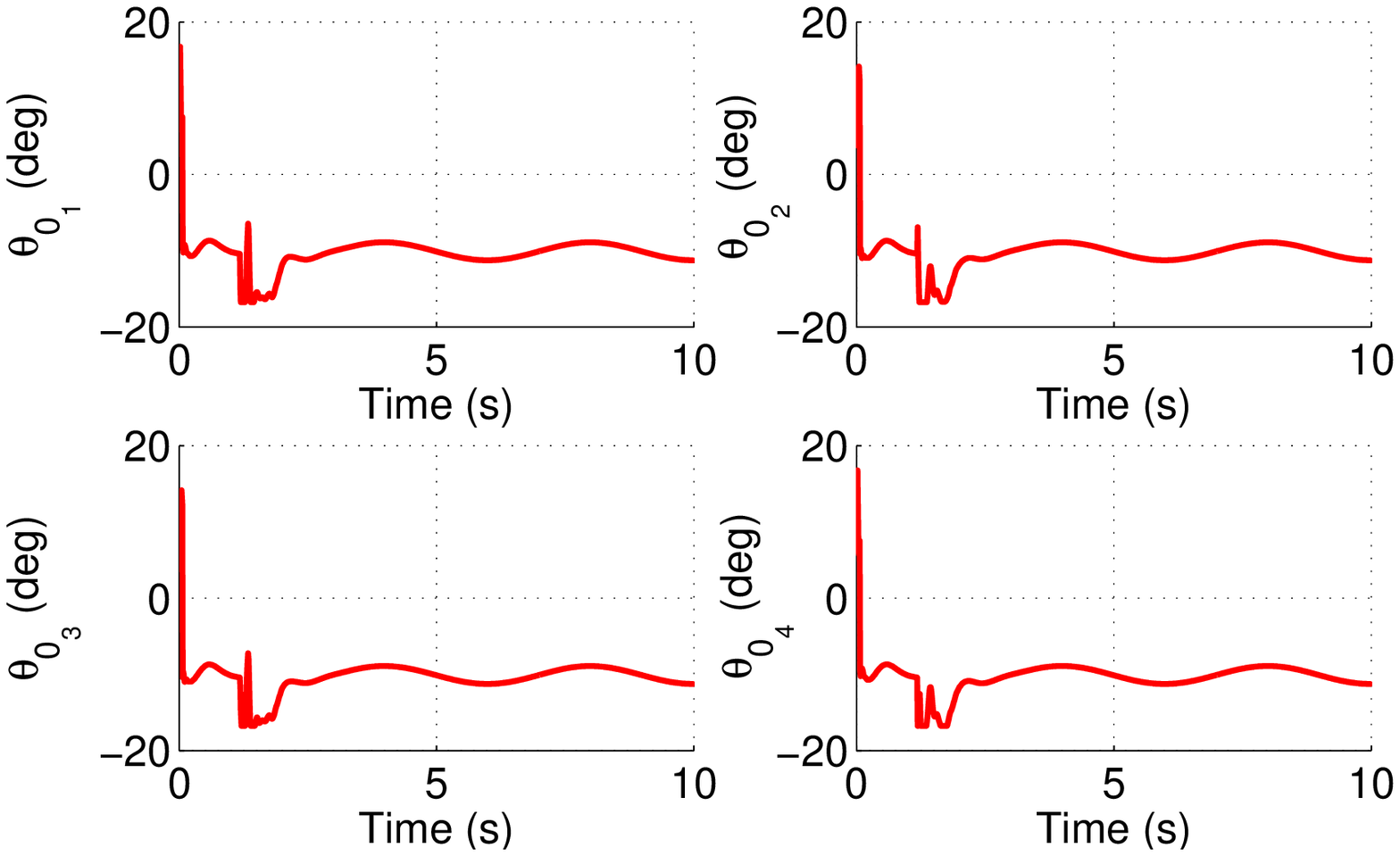}
	\caption{\textbf{Required values of Collective inputs variation for tracking trajectory during inverted flight}}
	\label{collective_inverted}
\end{figure}

The rapid changes in the commanded thrust from the individual rotors is shown in Fig. ~\ref{coeff_of_thrust_inverted}. Figure~\ref{collective_inverted} shows that all the rotors operate at negative collective input after the quadrotor is inverted to produce thrust in upward direction for tracking the trajectory in upside down attitude.

\begin{figure}[ht]
	\centering
	\includegraphics[width = 0.5\textwidth]{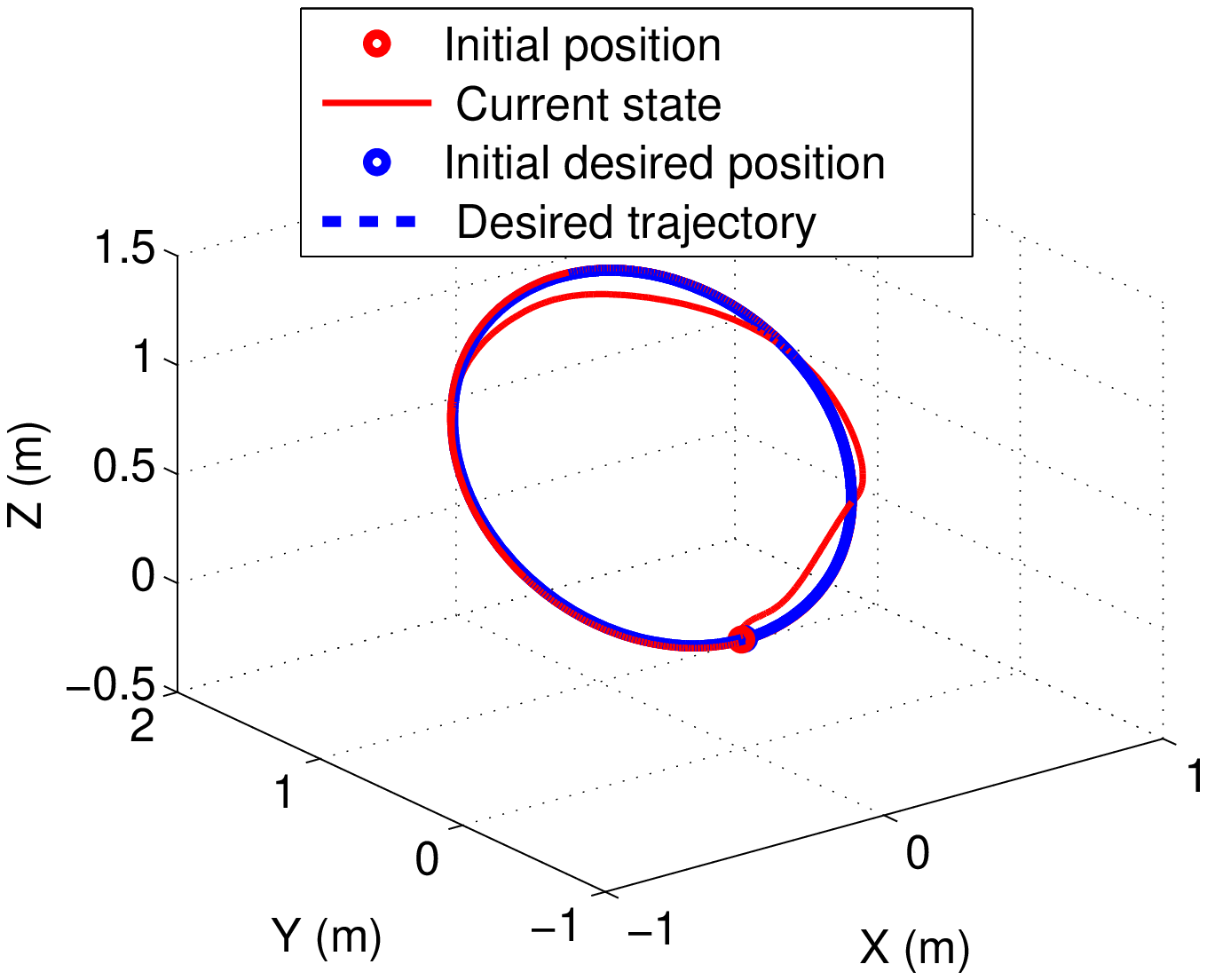}
	\caption{\textbf{Three Dimensional Variation in Position in tracking trajectory during inverted flight}}
	\label{3d_inverted}
\end{figure}

\section{Conclusions}\label{sec:conclusions}
This paper discusses the development of the flight dynamics model of a variable-pitch quadrotor which is suitable for model based controller design. The thrust and moment for each rotor is calculated using Blade Element Theory and momentum theory. Due to its ability to generate negative thrust, the variable-pitch quadrotor is known to offer higher controller bandwidth, which is suitable for aggressive maneuvering and inverted flight. A novel nonlinear controller is developed using dynamic inversion approach and demonstrated for stabilization, tracking, flipping and inverted flying of the variable pitch quadrotor. The challenge associated with control allocation, due to non-rational relation between blade pitch angle and rotor propulsive forces, is solved using an additional loop in the control design. The strategy of controlling the quadrotor by changing the blade pitch angle is validated by showing attitude stabilization in real flight for a variable pitch quadrotor.
The change in coordinate system due to flipping is taken care by introducing suitable variable for booking keeping of the orientation. The performance of controller is demonstrated through numerical simulations. As the controller is derived using six-DOF model, it is generic and can be employed for the whole flight regime.


\bibliographystyle{vancouver}
\bibliography{quadRefs}

\end{document}